\pdfoutput=1

\documentclass[letterpaper,11pt]{article}
\usepackage{cite}
\usepackage{amsmath,amssymb,amsfonts,array}
\usepackage[multiple,stable]{footmisc}
\usepackage[amsmath,amsthm,thmmarks]{ntheorem}
\allowdisplaybreaks[4]
\usepackage[noconfig]{refstyle}
\usepackage[dvipsnames]{xcolor}
\usepackage{array,tabularx,booktabs,geometry,subfigure,graphicx,longtable}
\usepackage[bf]{caption}
\usepackage{tikz}
\usetikzlibrary{shapes,arrows}
\tikzstyle{block} = [rectangle, draw, text width=7em, text centered, rounded corners, minimum height=3em]
\usepackage{color}
\usepackage[hyperfootnotes=true, linktocpage=true, colorlinks, citecolor=blue, linkcolor=blue, urlcolor=Maroon, filecolor=Maroon]{hyperref}

\newtheorem{theorem}{Theorem}[section]

\newtheorem{definition}[theorem]{Definition}

\let\eqref=\relax
\newref{eq}{name={eq.~},Name={Eq.~},names={eqs.~},Names={Eqs.~},rngtxt={-},refcmd=(\ref{#1})}
\newref{f}{name={footnote~},Name={Footnote~},names={footnotes~},Names={Footnotes~}}
\newref{app}{name={appendix~},Name={Appendix~},names={appendixes~},Names={Appendixes~}}
\newref{tab}{name={table~},Name={Table~},names={tables~},Names={Tables~}}
\newref{ch}{name={chapter~},Name={Chapter~},names={chapters~},Names={Chapters~}}
\newref{sec}{name={section~},Name={Section~},names={sections~},Names={Sections~}}
\newref{fig}{name={figure~},Name={Figure~},names={figures~},Names={Figures~}}
\numberwithin{equation}{section}

\newcommand{\be}{\begin{equation}}
\newcommand{\ee}{\end{equation}}
\newcommand{\bea}{\begin{equation}\begin{aligned}}	
\newcommand{\eea}{\end{aligned}\end{equation}}		

\newcommand{\iddots}{\mathinner{\mkern2mu\raise1pt\hbox{.}\mkern2mu \raise4pt\hbox{.}\mkern2mu\raise7pt\hbox{.}\mkern1mu}}

\providecommand{\id}{\leavevmode\hbox{\small$\mathrm{1}$\kern-3.8pt\normalsize$\mathrm{1}$}}

\newcommand{\beq}{\begin{equation}}
\newcommand{\eeq}{\end{equation}}
\newcommand{\ba}{\begin{array}}
\newcommand{\ea}{\end{array}}
\newcommand{\bean}{\begin{eqnarray*}}
\newcommand{\eean}{\end{eqnarray*}}
\newcommand{\eref}[1]{(\ref{#1})}

\newcommand{\comment}[1]{}

\begin{document}

\vspace{1cm}

\title{
       \vskip 40pt
       {\Large \bf Algebroids, Heterotic Moduli Spaces and the Strominger System}}

\vspace{2cm}

\author{
Lara~B.~Anderson,
James~Gray and
Eric~Sharpe
}
\date{}
\maketitle
\begin{center} {\small ${}^1${\it Department of Physics \\
Robeson Hall, 0435 \\ Virginia Tech\\ Blacksburg, VA  24061}}\\
 {\tt lara.anderson@vt.edu}, {\tt jamesgray@vt.edu}, {\tt ersharpe@vt.edu},

\end{center}

\begin{abstract}
\noindent
In this paper we study compactifications of heterotic string theory on manifolds satisfying the $\partial \overline{\partial}$-lemma. We consider the Strominger system description of the low energy supergravity to first order in $\alpha'$ and show that the moduli of such compactifications are subspaces of familiar cohomology groups such as $H^1(TX)$, $H^1(TX^{\vee})$, $H^1(\textnormal{End}_0(V))$ and $H^1(\textnormal{End}_0(TX))$. These groups encode the complex structure, K\"ahler moduli, bundle moduli and perturbations of the spin connection respectively in the case of a Calabi-Yau compactification. We investigate the fluctuations of only a subset of the conditions of the Strominger system (expected to correspond physically to F-term constraints in the effective theory). The full physical moduli space is, therefore, given by a further restriction on these degrees of freedom which we discuss but do not explicitly provide. This paper is complementary to a previous tree-level worldsheet analysis of such moduli and agrees with that discussion in the limit of vanishing $\alpha'$. The structure we present can be interpreted in terms of recent work in Atiyah and Courant algebroids, and we conjecture links with aspects of Hitchin's generalized geometry to heterotic moduli. 
\end{abstract}

\thispagestyle{empty}
\setcounter{page}{0}
\newpage

\tableofcontents

\section{Introduction}\label{intro}

In a seminal paper, Strominger wrote down the conditions for the most general geometric compactification of heterotic string theory which gives rise to an ${\cal N}=1$ theory with a maximally symmetric ${\cal N}=1$ vacuum  \cite{Strominger:1986uh}. A question of principle importance for the use of these spaces in string phenomenology is the nature of their moduli space. The massless charged matter of the theory is required, in a realistic compactification, to be compatible with our experimental observations. The uncharged moduli of the theory must be fixed, or stabilized, at vevs resulting in realistic values for the gauge and Yukawa couplings.

In the special case of a Calabi-Yau compactification, it is well known how to compute the field content of the low energy theory. 
Naively, the moduli of such a heterotic solution are the metric and bundle
moduli, which, via theorems by Yau, Donaldson, and Uhlenbeck and Yau, are counted by the dimensions of the cohomology groups $H^1(TX)$, $H^1(TX^{\vee})$ and $H^1(\textnormal{End}_0(V))$. 
Even in this case, there are some subtleties in counting the number of massless fields due to the gauge field structure in the problem, as has recently been discussed in \cite{Anderson:2010mh,Anderson:2011ty,Anderson:2013qca}.

In addition to Calabi-Yau compactifications, there has also
been considerable interest in non-K\"ahler solutions of the
heterotic string, see for example \cite{ms,Fernandez:2008wa,Grana:2004bg,Grana:2005sn,Fu:2008zh,Adams:2006kb,Adams:2009zg,Adams:2009av,Adams:2009tt,Blaszczyk:2011ib,Quigley:2011pv,Quigley:2012gq,Melnikov:2012nm,Fu:2006vj,Dasgupta:1999ss,Goldstein:2002pg,Becker:2006et,Maxfield:2013wka,Andriot:2011iw,Hohm:2011ex,Heckman:2013sfa,Becker:2009df,Becker:2004ii,Becker:2003yv,Gray:2012md,Klaput:2013nla,Klaput:2012vv,Klaput:2011mz,Lukas:2010mf,Gurrieri:2007jg,Gurrieri:2004dt,LopesCardoso:2002hd,Gauntlett:2003cy,LopesCardoso:2003af,Lechtenfeld:2010dr,Chatzistavrakidis:2009mh,Chatzistavrakidis:2008ii}. 
In particular, as Yau's theorem can no longer be applied to relate metric
moduli to deformations of the complex and K\"ahler structure, a longstanding problem in non-K\"ahler
heterotic compactifications has been to understand their massless degrees of freedom.

A proposal for the physical moduli of non-K\"ahler
solutions to the Strominger system was given in \cite{ms},  
at zeroth order in $\alpha'$ (see {\it e.g.}
\cite{Tseng:2011gv,Cyrier:2006pp} for other related work).
The analysis revolved around computing supersymmetric marginal operators in the (0,2) supersymmetric worldsheet theory.  
Such worldsheet-based computations are only valid for backgrounds associated to weakly coupled nonlinear sigma models, and in non-K\"ahler heterotic compactifications, the existence of such a weakly coupled regime is not always clear.  
Nevertheless, such a worldsheet-based analysis is relevant for those cases where the volume is large compared to the string scale. Given the hierarchy of scales which is seen in nature, this may well be the types of compactification of interest in heterotic string phenomenology. In addition, such work might act as a starting point for a more general discussion. 

In this paper, we return to the question of moduli in non-K\"ahler solutions to the Strominger system, utilizing the complementary approach of studying the low-energy supergravity.  We derive results valid through first order in $\alpha'$, a technical improvement over the discussion of \cite{ms}, with our results being in agreement with that work in the limit where $\alpha' \to 0$.  On the other hand, unlike \cite{ms}, we `overcount' the massless degrees of freedom of the compactification as, as we will describe, we shall only analyze a subset of the conditions imposed upon the system by Strominger. Our supergravity deformation analysis will also be restricted to compactifications on spaces satisfying the 
$\partial \overline{\partial}$-lemma, which can be stated as follows:

\vspace{0.2cm}

{\it {\bf Lemma:} Let $X$ be a compact K\"ahler manifold. For $A$ a $d$-closed $(p,q)$-form, the following statements are equivalent.
\begin{eqnarray} \label{ddbarlemma}
A = \overline{\partial} C  \Leftrightarrow A = \partial C' \Leftrightarrow A = d C'' \Leftrightarrow A= \partial \overline{\partial} \tilde{C} \Leftrightarrow A = \partial \hat{C} + \overline{\partial} \check{C}
\end{eqnarray}
 for some $C, C', C'', \tilde{C}$ and $\check{C}$.

\vspace{0.1cm}

We define a $\partial \overline{\partial}$-manifold to be any manifold, K\"ahler or not, which satisfies this lemma.}

\vspace{0.2cm}

There are many known examples of Strominger compactifications on spaces which obey the $\partial \overline{\partial}$-lemma. These include standard and non-standard embeddings on Calabi-Yau threefolds, as well as ``fully non-K\"ahler" possibilities \cite{Fu:2008zh}. It is known that the $\partial \overline{\partial}$-lemma holds for manifolds in the class ${\cal C}$ of Fujiki \cite{deligne,fujiki}. That is for manifolds which are compact and the meromorphic image of a compact K\"ahler space (which is not necessarily K\"ahler). It is also interesting to note that requiring the $\partial \overline{\partial}$-lemma leads to balanced structure of a metric being stable under small deformations (see \cite{fydg} for a recent discussion).

We will describe how the metric, spin connection, and bundle
deformations of such examples of the Strominger system are described as subspaces of ordinary bundle valued Dolbeault cohomology groups. These subspaces are given by nested kernels of maps between familiar cohomologies, such as $H^1(TX)$ and $H^2(\textnormal{End}_0 (V))$. The maps involved are defined by the geometric data of the compactification. Equivalently, the deformations are described by the first cohomology of a bundle which is not simply $TX\oplus TX^{\vee}\oplus \textnormal{End}_0(V)$ but rather a non-trivial merging of these components (and $H^1(\textnormal{End}_0(TX))$).  Our results can be expressed in the structure of Atiyah and Courant algebroids, and have tantalizing connections with Hitchin's generalized geometry.

\vspace{0.3cm}

In order to explain the structure we will present, it is helpful to make a comparison to a case which is already well known in the literature - that of the Atiyah class stabilization of complex structure moduli in Calabi-Yau threefold compactifications of heterotic theories \cite{Anderson:2010mh,Anderson:2011ty,Anderson:2013qca,atiyah}. Gauge fields in such a compactification must obey the Hermitian Yang-Mills equation at zero slope:
\begin{align} \label{mdela1}
F_{\overline{a}\overline{b}}&=0\;, \\ \label{mdela2}
g^{a\overline{b}} F_{a\overline{b}}&=0\;.
\end{align}
The first equation does not depend on the metric, but merely the complex structure and gauge bundle of the system. It states that the gauge bundle must be holomorphic. The second equation, which does depend upon the metric, states that the gauge bundle is poly-stable and of zero slope (in the Mumford sense). 

Since equation~(\ref{mdela1}) depends explicitly on the complex structure of the compactification, it is of no surprise that it leads to a stabilization of some of the complex structure moduli of the base Calabi-Yau threefold. The unstabilized complex structure moduli can be described as the kernel of the map,
\begin{eqnarray} \label{intromap}
H^1(TX) \stackrel{[F]}{\longrightarrow} H^2(\textnormal{End}_0(V))\;.
\end{eqnarray}
The map, as indicated in (\ref{intromap}), is determined by the cohomology class of the field strength of the gauge connection. The crucial point is that the moduli of the system can still be described in terms of a subspace of the ordinary bundle valued Dolbeault cohomologies, despite the extra structure in the system due to the gauge bundle. The constraint (\ref{intromap}), due to the equation (\ref{mdela1}), can be reproduced in the four dimensional theory by F-term constraints \cite{Witten:1985bz}. The constraints imposed by the condition (\ref{mdela2}), on the other hand, are reproduced by D-terms \cite{Witten:1985bz}. The D-term structure is, of course, entirely determined by the charges of the matter content of the theory.

Let us compare this discussion to the case described in the present paper. The most pertinent equations can be written as follows:
\begin{align} \label{thisone1}
F_{\overline{a} \overline{b}} &=0 \;, &&\;\;H = \frac{i}{2} (\overline{\partial}-\partial) J \;, \\ \label{thisone2}
g^{a\overline{b}}F_{a\overline{b}} &=0 \;,&&\;\;
g^{\overline{b}c} H_{\overline{b}c\overline{a}} = - 6 \overline{\partial}_{\overline{a}} \phi\;.
\end{align}

As with the Atiyah discussion, first two equations are again the focus of our discussion. The first of these is in fact, once more, the condition for bundle holomorphy, and as such we expect the Atiyah kernel (\ref{intromap}) to be part of the result. We find that the fluctuations consistent with the  other equation in (\ref{thisone1}) and the heterotic Bianchi identity are also subgroups of the Dolbeault cohomology groups with the maps determined by the field configurations.

We will show that the fluctuations in the metric, spin connection, and bundle moduli of this system are a subspace of $H^1(TX) \oplus H^1(TX^{\vee}) \oplus H^1(\textnormal{End}_0(V)) \oplus H^1(\textnormal{End}_0(TX))$. In a Calabi-Yau compactification the first three of these would be referred to as the complex structure, K\"ahler and bundle moduli respectively (whereas the last piece corresponds to a redundant description of the perturbations of the spin connection). In view of the different situation being considered here, we will refer to $H^1(TX^{\vee})$ as being associated to the ``Hermitian" rather than ``K\"ahler" moduli.  We will show that the unstabilized deformations consistent with (\ref{thisone1}) are given by $H^1({\cal H})$, described in the following form
\begin{eqnarray} 
H^1({\cal H}) = \left\{ \begin{array}{c} \textnormal{ker} (H^1({\cal Q}) \to H^2(TX^{\vee})) \\ \oplus \\  H^1(TX^{\vee}) \;,\end{array}\right.
\end{eqnarray}
where
\begin{eqnarray} 
H^1({\cal Q}) &=& \left\{\begin{array}{c}  H^1(\textnormal{End}_0(V) ) \oplus H^1(\textnormal{End}_0(TX))\\ \oplus \\ \label{h0Q} \textnormal{ker}( H^1(TX) \to H^2(\textnormal{End}_0(V)) \oplus H^2(\textnormal{End}_0(TX)))\;.\end{array}\right. 
\end{eqnarray}
These results subsume the Atiyah stabilization, and the maps are once again defined by the background fields of the solution to the Strominger system under consideration (see Section \ref{Sec:seq} for details). As we have mentioned, as described in the equations above, $H^1({\cal H})$ is a subgroup not of $H^1(TX) \oplus H^1(TX^{\vee}) \oplus H^1(\textnormal{End}_0(V))$ but rather $H^1(TX) \oplus H^1(TX^{\vee}) \oplus H^1(\textnormal{End}_0(V)) \oplus H^1(\textnormal{End}_0(TX))$. The extra contribution, in $H^1(\textnormal{End}_0(TX))$, corresponds to perturbations of the spin connection. In a Calabi-Yau compactification, for example, these degrees of freedom are redundant with the metric moduli. It turns out to be simpler in the analysis to treat these perturbations as separate from those of the metric, however, and their presence helps in linking the structure to the mathematics of Courant algebroids - hence their appearance here. 

As in the case of the Atiyah stabilization we expect the equations (\ref{thisone2}) to be encoded in terms of D-terms in the four dimensional theory. We will not however address these further restrictions in this paper. The crucial point is once more that, despite all of the extra structure of the Strominger system, the moduli are once again subspaces of the same cohomology groups that are familiar from Calabi-Yau compactifications, in the case where the compactification obeys the $\partial \overline{\partial}$-lemma.

\vspace{0.1cm}

We would like to highlight that, as we were finishing this work, we were made aware of closely related work that will appear concurrently with this paper \cite{svanes-ossa}.

\vspace{0.1cm}

The rest of this paper is structured as follows. In Section \ref{Sec:perturb} we perform a field theory perturbation analysis of the Strominger system. In Section \ref{Sec:seq} we relate this field theory discussion to a cohomological description, as described above. In Section \ref{Sec:algebroids} we describe the relationship of this analysis to previous work \cite{Baraglia:2013wua,garcia_ferndandez} relating the Strominger system to transitive Courant algebroids. We also discuss the relationship of the deformations we describe to that of a generalized complex structure on the total space of a specific bundle. In Section \ref{Sec:sigma} we describe the relationship of our work to previous research in the context of NLSM's. Finally, in Section \ref{discussion}, we conclude and discuss possible future directions of research. A technical appendix provides some details on the mathematics of Courant algebroids which are required in the text.


\section{Perturbing the Strominger system}\label{Sec:perturb}

Strominger has written down the conditions which are necessary and sufficient for a compactification of heterotic string theory to four dimensions to exhibit a maximally symmetric vacuum with ${\cal N}=1$ supersymmetry \cite{Strominger:1986uh}. They are,
\begin{itemize}
\item The compactification manifold must admit an integrable complex structure.
\item The fundamental form $J_{a \bar{b}}= i g_{a \bar{b}}$ must obey the following two equations\footnote{The curvature of the Levi-Civita connection appears 
in (\ref{bi1}) rather than a connection modified by terms involving $H$ because of the counting in $\alpha'$. The field strength $H$ is order $\alpha'$ 
thanks to the nature of the Bianchi identity and flux quantization conditions. Thus modifying the curvature that appears in (\ref{bi1}) by terms 
involving $H$ will only modify the discussion at second order in $\alpha'$. In this paper we will restrict ourselves to the first non-trivial order in the expansion.}:
\begin{align}  \label{bi1}
\partial \overline{\partial} J &= \frac{1}{30} i \alpha' \textnormal{Tr} F\wedge F-i \alpha' \textnormal{tr} R\wedge R \:, \\ \label{ddagger}
d^{\dagger} J &= i ( \partial - \overline{\partial}) \textnormal{ln} || \omega || \;.
\end{align}
In the above expression, $||\omega||$ is the norm of the holomorphic $(3,0)$ form associated to the $SU(3)$ structure admitted by the compactification manifold.
\item The Yang-Mills field strength must satisfy
\begin{align}
J^{a \bar{b}} F_{a\bar{b}} &=0 \:, \\
F_{ab} = F_{\bar{a} \bar{b}} &= 0\;.
\end{align}
\end{itemize}

It is useful to split up the above conditions by introducing new quantities in the form of the NS-NS field strength $H$ and the dilaton $\phi$. In addition to making certain parallels between the equations for $J$ and $F$ manifest, this will be useful when we come to consider the effects of flux quantization in this setting. 

Following Strominger, we define $H$, via the expression
\begin{align}  \label{heq}
H = \frac{i}{2} ( \overline{\partial} - \partial) J \;,
\end{align}
and $\phi$, via
\begin{align} \label{phieq}
\phi = \frac{1}{8} \textnormal{ln} || \omega || + \phi_0 \:,
\end{align}
where $\phi_0$ is a constant.

Using these expressions we can rewrite some of the conditions of the Strominger system. From the Bianchi identity, equation (\ref{bi1}), we then have that
\begin{align}
dH &= - \frac{1}{30} \alpha' \textnormal{tr} F \wedge F + \alpha' \textnormal{tr} R \wedge R \;, \\ \label{mrH}
\Rightarrow H &= \tilde{H} - \frac{1}{30}\alpha' \omega_3^{\textnormal{YM}} + \alpha' \omega_3^{\textnormal{L}} \;,
\end{align}
where the difference in Chern Simons terms is well defined thanks to the anomaly cancelation condition. In this expression $d \tilde{H}=0$ and $\tilde{H}$ obeys an integer valued flux quantization condition.

 We can also rewrite the equation (\ref{ddagger}) in terms of $H$ and $\phi$, using (\ref{heq}) (this is also done, for example, in \cite{Gillard:2003jh}). We have, by definition, that
\begin{align}
(d^{\dagger}J)_{\overline{a}} &= \nabla^{b} J_{\overline{a} b}  \;,
\end{align}
whereas, from (\ref{heq}),
\begin{align}
H_{\overline{a} \overline{b} c} &= \frac{3}{2}i \overline{\partial}_{[\overline{a}} J_{\overline{b}c]} = \frac{3}{2} i \overline{\nabla}_{[\overline{a}} J_{\overline{b}c]} \;,\\ 
\Rightarrow H_{\overline{a} \overline{b} c} g^{c \overline{b}} &= \frac{3}{4}i \left( g^{c\overline{b}} \overline{\nabla}_{\overline{a}} J_{\overline{b}c} - g^{c \overline{b}} \overline{\nabla}_{\overline{b}} J_{\overline{a}c}\right) \\
&= -\frac{3}{4}i \nabla^c J_{\overline{a}c} \;.
\end{align}
Given this, we find the following:
\begin{align}
d^{\dagger}J &= i (\partial - \overline{\partial}) 8 \phi \;, \\
\Rightarrow \frac{4}{3}i H_{\overline{a}\overline{b}c}g^{c\overline{b}} &= - 8 i \overline{\partial}_{\overline{a}} \phi \;,\\ \label{second}
\Rightarrow H_{\overline{b}c\overline{a}}g^{\overline{b}c} &= -6 \overline{\partial}_{\overline{a}} \phi \;.
\end{align}
Similarly we find
\begin{align} \label{third}
H_{\overline{b}c a} g^{\overline{b}c} = 6 \partial_a \phi\;,
\end{align}
which is just the conjugate of (\ref{second}).

\vspace{0.1cm}

Thus, finally, we obtain the form we shall use for the Strominger equations. We have,
\begin{align} \label{mrbi}
dH &= - \frac{1}{30} \alpha' \textnormal{tr} F \wedge F + \alpha' \textnormal{tr} R \wedge R \;, \\ \label{first1}
H &= \frac{i}{2} ( \overline{\partial} - \partial) J \;, \\ \label{mrhol}
F_{ab}&=F_{\overline{a} \overline{b}} =0 \;,
\end{align}
as the equations involving no contractions with the metric, and
\begin{align} \label{d1}
H_{\overline{b}c\overline{a}}g^{\overline{b}c} = -6 \overline{\partial}_{\overline{a}} \phi \;\;\; ,\;\;\; H_{\overline{b}c a} g^{\overline{b}c} = 6 \partial_a \phi \;\;\; \textnormal{and} \;\;\; g^{a\overline{b}} F_{a \overline{b}} =0
\end{align}
for the remaining relations.

Our expectation is that the first set of equations above are the generalizations to the Strominger system of the Bianchi identity and holomorphy conditions of the Calabi-Yau case. As such the restrictions they enforce on the moduli space of the system, that we anticipate is the space of F-flat fluctuations, should be describable as kernels of maps between ordinary Dolbeault cohomologies. We will show that this is so, in this section and the next, for a manifold satisfying the $\partial \overline{\partial}$-lemma. The second set of equations is the generalization of poly-stability to the non-K\"ahler situation. As such we expect these conditions to be more involved, and case dependent, in their analysis. We shall not discuss these conditions in detail in this paper.

To understand how the equations (\ref{mrbi}-\ref{mrhol}) restrict the moduli space, we assume we have a solution to the system, and perturb the complex structure, the fundamental two form, the spin connection, the gauge fields and the Neveu-Schwarz two form about that configuration. We begin with equation (\ref{first1}).

\subsection{Perturbing equation (\ref{first1}) and flux quantization}

The expression for $H$, (\ref{first1}), as written, explicitly involves complex coordinates. In order to facilitate analysis of perturbations of this equation under fluctuations of the complex structure, we rewrite the equation in terms of real coordinates as follows.
\begin{align} \label{first}
H_{ijk} &= \frac{i}{2} \left( \Pi_{i}^{(+)l} \Pi_j^{(+)m} \Pi_{k}^{(-)n}+ \Pi_{i}^{(+)l} \Pi_j^{(-)m} \Pi_{k}^{(+)n}+ \Pi_{i}^{(-)l} \Pi_j^{(+)m} \Pi_{k}^{(+)n} \right)dJ_{lmn} \\ \nonumber
&- \frac{i}{2} \left( \Pi_{i}^{(-)l} \Pi_j^{(-)m} \Pi_{k}^{(+)n}+\Pi_{i}^{(-)l} \Pi_j^{(+)m} \Pi_{k}^{(-)n}+\Pi_{i}^{(+)l} \Pi_j^{(-)m} \Pi_{k}^{(-)n} \right) dJ_{lmn} \;,
\end{align}
where
\begin{equation}
 \Pi_i^{(\pm) j} = \frac{1}{2} \left( 1 \pm i J \right)_i^{\;j} 
\end{equation}
are the projectors onto holomorphic and anti-holomorphic coordinates.

The perturbation to $\tilde{H}$, as defined in equation (\ref{mrH}) also has to be of the form $d \delta B$. This is because the integral of $\tilde{H}$ over any three cycle is quantized. If there were a harmonic part to the perturbation in $\delta \tilde{H}$ (the other possibility) this would violate the integer quantization. Given this, we have,
 \begin{align} \label{realind}
3 d_{[i} \delta B_{jk]} -\frac{1}{30} \alpha' \delta \omega^{\textnormal{YM}}_{3ijk} + \alpha' \delta \omega^L_{3ijk} =\delta H_{ijk}\;,
\end{align}
where $H_{ijk}$ is given by equation (\ref{first}). 

We now analyze this equation, order by order in $\alpha'$.

\subsubsection{Zeroth order in $\alpha'$}

To understand the implications of equation (\ref{realind}) we now examine its components in terms of complex coordinates adapted to the original, unperturbed complex structure. Taking $i,j,k = \bar{a} ,\bar{b}, \bar{c}$ we find,
\begin{align} 
3 \overline{\partial}_{[\bar{a}} \delta B _{\bar{b} \bar{c}]} &= \frac{i}{2} \left( -\frac{i}{2}  \delta J_{\overline{c}}^{\; c} dJ_{\overline{a} \overline{b}c}-\frac{i}{2}  \delta J_{\overline{b}}^{\; b} dJ_{\overline{a} b \overline{c}}-\frac{i}{2}  \delta J_{\overline{a}}^{\; a} dJ_{a \overline{b} \overline{c}} \right) \;,\\
&= \frac{3}{4} \left( \delta J_{\overline{c}}^{\; d} d_{[\overline{a}} J_{\overline{b}c]}  + \delta J_{\overline{b}}^{\; d} d_{[\overline{a}} J_{c \overline{c}]}  + \delta J_{\overline{a}}^{\; d} d_{[c} J_{\overline{b}\overline{c}]}  \right) \delta^c_d \;, \\
&= 3 \delta J_{[\overline{c}}^{\;d} d_{\overline{a}} J_{\overline{b}c]} \delta_d^c \;,\\ 
&= 3 \delta J_{[\overline{c}}^{\;d} \nabla_{\overline{a}} J_{\overline{b}c]} \delta_d^c \;, \\
&=3   \nabla_{[\overline{a}} (\delta J_{\overline{c}}^{\;d}J_{\overline{b}c]}) \delta_d^c  \;,\\
&= \frac{3}{2} \nabla_{[\overline{a}} (\delta J_{\overline{c}}^{\;d}J_{\overline{b}]c}) \delta_d^c \;,\\ \label{allbar}
&= - \frac{3}{2}i \overline{\partial}_{[\overline{a}} \delta J_{\overline{c} \overline{b}]} \;.
\end{align}
In the above we have used that $\delta J \in H^1(TX)$ and, given that $J_{ij}$ is a $(1,1)$ form, 
\begin{align}
& ~~~~~J_i^{\;j}J_{jl} J_k^{\;l} = J_{ik} \;,\\ 
&\Rightarrow \delta J_i^{\;j} J_{jl}J_k^{\;l} + J_i^{\;j} \delta J_{jl} J_k^{\;l} + J_i^{\;j} J_{jl} \delta J_k^{\;l} = \delta J_{ik} \;, \\
&\Rightarrow \delta J_{\overline{c}\overline{b}} = \delta J_{\overline{c}}^{\;c} J_{c\overline{b}} (-i) + (-i) \delta J_{\overline{c}\overline{b}} (-i) + (-i) J_{\overline{c}c} \delta J_{\overline{b}}^{\;c} \;, \\ \label{mrtheident}
&\Rightarrow \delta J_{\overline{c} \overline{b}} = i \delta J_{[\overline{c}}^{\;c} J_{\overline{b}] c} \;.
\end{align}

Given (\ref{allbar}), any change in $(0,2)$ component of $J$ can be compensated for by an appropriate change in $B$ (which therefore becomes part of the reduction ansatz):
\begin{align} \label{Bcompensator}
\delta B_{\bar{b}\bar{c}} = \frac{1}{2} i \delta J_{\bar{b} \bar{c}} + \delta B'_{\bar{b}\bar{c}} \;.
\end{align}
Here $\delta B'$ is an arbitrary $\overline{\partial}$ closed $(0,2)$ form. 

\vspace{0.1cm}

Returning to (\ref{realind}) we now consider the remaining possibility for the components (up to conjugation), $i,j,k=\bar{a},\bar{b},c$.
\begin{align}
3 d_{[\bar{a}} \delta B_{\bar{b}c]} &= \frac{i}{2} (d \delta J)_{\overline{a} \overline{b} c} + \frac{i}{2} \left(  \frac{i}{2} \delta J_{\overline{a}}^{\;d} dJ_{d \overline{b}c} + \frac{i}{2} \delta J_{\overline{b}}^{\;d} dJ_{\overline{a}d c} + \frac{i}{2} \delta J_{\overline{a}}^{\;d} dJ_{d \overline{b}c} + \frac{i}{2} \delta J_{\overline{b}}^{\;d} dJ_{\overline{a}dc} \right) \;, \\ 
&=  \frac{i}{2} (d \delta J)_{\overline{a} \overline{b} c} -\frac{1}{2} \delta J_{[\overline{a}}^{\;d} dJ_{\overline{b}]cd} - \frac{1}{2} \delta J_{[\overline{a}}^{\;d} dJ_{\overline{b}]cd} \;, \\
&=  i \overline{\partial}_{[\overline{a}} \delta J_{\overline{b}]c} + \frac{i}{2} \partial_c \delta J_{\overline{a}\overline{b}} - \delta J_{[\overline{a}}^{\;d} dJ_{\overline{b}]cd} \;.
\end{align}
Expanding out the left hand side as well we obtain 
\begin{align}
2 \overline{\partial}_{[\overline{a}} \delta B_{\overline{b}]c} + \partial_c \delta B_{\overline{a} \overline{b}}&=  i  \overline{\partial}_{[\overline{a}} \delta J_{\overline{b}]c} + \frac{1}{2}i \partial_c \delta J_{\overline{a}\overline{b}}- \delta J_{[\overline{a}}^{\;d} dJ_{\overline{b}]cd} \;.
\end{align}
Now we use our solution (\ref{Bcompensator}) in the previous equation:
\begin{align} \label{penul}
\Rightarrow 2 \overline{\partial}_{[\bar{a}} \delta B_{\bar{b}]c} + \partial_c \delta B'_{\bar{a}\bar{b}} &=- \delta J_{[\bar{a}}^d \partial J_{\bar{b}]cd} +i \overline{\partial}_{[\bar{a}} \delta J_{\bar{b}]c} \;.\end{align}

As stated earlier, we will consider manifolds satisfying the $\partial \overline{\partial}$-lemma (\ref{ddbarlemma}). Given that $\delta B'$ is $\overline{\partial}$ closed, we see that $\partial_c \delta B_{\overline{a}\overline{b}}'$ is $d$ closed and thus, by the lemma, $\partial_c \delta B_{\overline{a}\overline{b}}'= \overline{\partial}_{[\overline{a}} \Lambda_{\overline{b}]c}$ for some $(1,1)$ form $\Lambda$. We then find that (\ref{penul}) gives us the following:
\begin{align} \label{themap}
\delta J_{[\bar{a}}^d \partial J_{\bar{b}]cd} &= i \overline{\partial}_{[\bar{a}} \delta J_{\bar{b}]c} - 2 \overline{\partial}_{[\bar{a}} \delta B_{\bar{b}]c} -\overline{\partial}_{[\overline{a}} \Lambda_{\overline{b}]c} \;.
\end{align} 
This is the form of the fluctuation equation, to zeroth order in $\alpha'$, that we will require for the rest of the paper.

\subsubsection{First order in $\alpha'$}

We now wish to add the first order Chern-Simons terms back into our analysis of (\ref{realind}). For this we need to know the variation of a Chern-Simons term in an appropriate form: 
\begin{align}\label{csvar}
\frac{1}{3!} \omega^{\textnormal{YM}}_{3 ijk} &=  \delta _{xy} A^x_{[i} \partial_j A^y_{k]} + \frac{2}{3} f_{xyz} A^{x}_{[i} A^y_j A^z_{k]} \;. \\ 
\Rightarrow \frac{1}{3!} \delta \omega^{\textnormal{YM}}_{3 ijk} &=  \delta_{xy} \delta A^x_{[i} \partial_j A^y_{k]} + \delta_{xy} A^x_{[i} \partial_j \delta A^y_{k]} + 2 f_{xyz} \delta A^x_{[i} A^y_j A^z_{k]} \;, \\ 
&= \delta_{xy} \delta A^x_{[i} \partial_j A^y_{k]} +\delta_{xy} \partial_{[j} \left( A^x_i \delta A^y_{k]} \right) - \delta_{xy} \partial_{[j} ( A^x_i )\delta A^y_{k]} +2 f_{xyz} \delta A^x_{[i} A^y_j A^z_{k]} 
\;, \\ \label{var1}
&=  \delta_{xy} \partial_{[i} \left( \delta A^y_j A^x_{k]}\right) + \delta_{xy} \delta A^x_{[i} F^y_{jk]} \;.
\end{align}
Naturally, we have similar expressions for $\omega_3^L$. Taking $W$ to be the spin connection we have  
\begin{align} \label{cslvar}
\frac{1}{3!} \omega^{\textnormal{L}}_{3 ijk} &= W^{\alpha \beta}_{[i} \partial_j W^{\beta \alpha}_{k]} + \frac{2}{3} W^{\alpha \beta}_{[i} W^{\beta \gamma}_j W^{\gamma \alpha}_{k]} \;, \\  \label{var2} 
\Rightarrow \frac{1}{3!} \delta \omega^{\textnormal{L}}_{3 ijk} &= \partial_{[i} \left( \delta W^{\alpha \beta}_j W^{\beta \alpha}_{k]}\right) + \delta W^{\alpha \beta}_{[i} R^{\beta \alpha}_{jk]} \;.
\end{align}

With these expressions in hand we can return to (\ref{realind}). Consider the $\bar{a}\bar{b}\bar{c}$ component below:
\begin{align}
3 \overline{\partial}_{[\bar{a}} \delta B_{\overline{b}\overline{c}]} - \frac{2}{10} \alpha' \left( \overline{\partial}_{[\overline{a}} \left( \delta A^y_{\overline{b}} A^x_{\overline{c}]} \delta_{xy} \right)\right) +  6 \alpha' \left( \overline{\partial}_{[\overline{a}} \left( \delta W^{\alpha \beta}_{\overline{b}} W^{\beta \alpha}_{\overline{c}]}\right)\right) = - \frac{3}{2} i \overline{\partial}_{[\bar{a}} \delta J_{\bar{c}\bar{b}]} \;.
\end{align}
In the above we have used the vanishing of the $(0,2)$ component of the background field strength and curvature two form\footnote{One easy way to see that the $(0,2)$ component of the curvature two form vanishes is via the $\alpha'$ expansion and the relation of the Levi-Civita and Chern connections. It is well known that the curvature of the $H$ deformed connection is a $(1,1)$ form (see \cite{Fernandez:2008wa} for an example of a discussion of this in the current context). The curvature of the Levi-Civita connection is the same as this at zeroth order in $\alpha'$. Therefore, in this term which is already order $\alpha'$ we can take the curvature two form to be zero while working to linear order.}. This leads to the generalization of (\ref{Bcompensator})
\begin{align}
\delta B_{\bar{b}\bar{c}} = \frac{2}{30} \alpha' \left( \delta A^y_{[\overline{b}} A^x_{\overline{c}]} \delta_{xy} \right) -2 \alpha'\left( \delta W^{\alpha \beta}_{[\overline{b}} W^{\beta \alpha}_{\overline{c}]}\right)+\frac{i}{2} \delta J_{\bar{b} \bar{c}} + \delta B'_{\bar{b}\bar{c}} \;,
\end{align}
where $\delta B'$ is a $\overline{\partial}$ closed form. The only other component of (\ref{realind}), up to conjugation, is the ${\overline{a} \overline{b} c}$ one. Making use of our previous analysis for the zeroth order pieces, and equations (\ref{var1}) and (\ref{var2}), we find the following:
\begin{align}
&2 \overline{\partial}_{[\bar{a}} \delta B_{\bar{b}]c} + \partial_c \delta B_{\bar{a}\bar{b}}  - \alpha' \frac{1}{30} \delta \omega^{\textnormal{YM}}_{3 \overline{a} \overline{b} c} + \alpha' \delta \omega^{\textnormal{L}}_{3 \overline{a} \overline{b} c}\\ \nonumber 
= &~~i \overline{\partial}_{[\bar{a}} \delta J_{\bar{b}]c}- \delta J_{[\bar{a}}^{\;d} \partial J_{\bar{b}]cd}  +\frac{1}{2}i \partial_c \delta J_{\overline{a} \overline{b}} \;, \\ 
\Rightarrow&~~   \delta J_{[\overline{a}}^{\;d}  \partial J_{\overline{b}]cd} + \partial_c\frac{2}{30} \alpha' \left( \delta A^y_{[\overline{a}} A^x_{\overline{b}]} \delta_{xy} \right) - 2 \alpha' \partial_c \left( \delta W^{\alpha \beta}_{[\overline{a}} W^{\beta \alpha}_{\overline{b}]}\right) - \frac{1}{30} \alpha' \delta \omega^{\textnormal{YM}}_{3 \overline{a} \overline{b} c} + \alpha' \delta \omega^{\textnormal{L}}_{3 \overline{a} \overline{b} c} \;, \\ 
\nonumber 
&=~~ -2\overline{\partial}_{[\overline{a}} \delta B_{\overline{b}] c} + i \overline{\partial}_{[\overline{a}} \delta J_{\overline{b}]c} - \partial_{c} \delta B'_{\overline{a} \overline{b}} \;, \\
\Rightarrow&~~ \delta J_{[\overline{a}}^{\;d}  \partial J_{\overline{b}]cd} -\frac{2}{30} \alpha' \overline{\partial}_{[\overline{a}} \left( \delta A^y_{\overline{b}]} A^x_c \delta_{xy}\right) +\frac{2}{30} \alpha' \overline{\partial}_{[\overline{a}} \left(  A^y_{\overline{b}]} \delta A^x_c \delta_{xy}\right) -\frac{4}{30} \alpha' \delta_{xy} \delta A^x_{[\overline{a}} F^y_{\overline{b}]c} \\ \nonumber
& ~~ + 2\alpha' \overline{\partial}_{[\overline{a}} \left( \delta W^{\alpha \beta}_{\overline{b}]} W^{\beta \alpha} _c\right)-2  \alpha' \overline{\partial}_{[\overline{a}} \left(  W^{\alpha \beta}_{\overline{b}]} \delta W^{\beta \alpha} _c\right) + 4 \alpha' \delta W^{\alpha \beta}_{[\overline{a}} R^{\beta \alpha}_{\overline{b}]c} \\ \nonumber &= -2\overline{\partial}_{[\overline{a}} \delta B_{\overline{b}] c} + i \overline{\partial}_{[\overline{a}} \delta J_{\overline{b}]c} - \partial_{c} \delta B'_{\overline{a} \overline{b}} \;, \\ 
\Rightarrow&~~ \delta J_{[\overline{a}}^{\;d}  \partial J_{\overline{b}]cd} -\frac{4}{30} \alpha' \delta_{xy} \delta A^x_{[\overline{a}} F^y_{\overline{b}]c} + 4 \alpha' \delta W^{\alpha \beta}_{[\overline{a}} R^{\beta \alpha}_{\overline{b}]c} =    i \overline{\partial}_{[\overline{a}} \delta J_{\overline{b}]c}-2\overline{\partial}_{[\overline{a}} \delta B_{\overline{b}] c}  - \overline{\partial}_{[\overline{a}} \Lambda^{\alpha'}_{\overline{b}]c} \;.
\end{align}
Here $\Lambda^{\alpha'}$ is the order $\alpha'$ corrected version of the $(1,1)$ form $\Lambda$ seen in the zeroth order result and we have once again made use of the $\partial \overline{\partial}$-lemma. This is the form of the perturbation equation (\ref{themap}), corrected to order $\alpha'$, that we will require in the rest of the paper.

\subsection{Overview of the F-flat conditions from the Strominger system}

The above analysis in fact completes our field theoretic discussion of the fluctuations of the Bianchi identity and ``F-term" relations (\ref{mrbi}-\ref{mrhol}). That the fluctuations are compatible with the Bianchi identity (\ref{mrbi}) is guaranteed by the form of $H$ which we perturbed, (\ref{mrH}). The constraint of bundle holomorphy, equation (\ref{mrhol}), has already be analyzed, in the fashion being discussed here, in the literature \cite{Anderson:2010mh,Anderson:2011ty,Anderson:2013qca}. In addition, it will be useful in what follows to add the equation describing how the holomorphic tangent bundle remains holomorphic under deformations.

Combining these results we have the following constraints on the fluctuations of the Strominger system on a manifold satisfying the $\partial \overline{\partial}$-lemma:
\begin{align}\label{one}
\delta J_{[\overline{a}}^{\;d}  \partial J_{\overline{b}]cd} -\frac{4}{30} \alpha' \delta_{xy} \delta A^x_{[\overline{a}} F^y_{\overline{b}]c} + 4 \alpha' \delta W^{\alpha \beta}_{[\overline{a}} R^{\beta \alpha}_{\overline{b}]c} &=    i \overline{\partial}_{[\overline{a}} \delta J_{\overline{b}]c}-2\overline{\partial}_{[\overline{a}} \delta B_{\overline{b}] c}  - \overline{\partial}_{[\overline{a}} \Lambda^{\alpha'}_{\overline{b}]c} \;,\\ \label{two}
i\delta J_{[\overline{a}}^{\;d} F_{\overline{b}]d} &=  2 D_{[\overline{a}} \delta A_{\overline{b}]} \;,  \\ \label{three}
i \delta J_{[\overline{a}}^{\;d} \hat{R}_{\overline{b}]d} &= 2 \hat{\nabla}_{[\overline{a}} \delta \hat{W}_{\overline{b}]} \;.
\end{align}
Note that the last of these equations contains hatted quantities which refer to curvatures, derivatives and perturbations associated with the Chern connection. The content of (\ref{three}) is simply the well known fact \cite{huyb} that the curvature of the Chern connection of a holomorphic tangent bundle is a $(1,1)$ form (and can remain so under perturbation). We will only require the zeroth order result in our analysis of the Strominger system to first order in $\alpha'$ because of the explicit $\alpha'$ factors appearing in (\ref{one}). In such a situation, one may remove the hats from equation (\ref{three}) and everywhere replace the Chern with the Levi-Civita connection (which it reduces to at zeroth order in $\alpha'$).

In the next section we will show how these constraints can be understood in cohomological terms. More precisely, we will show that fluctuations  satisfying (\ref{one}) and (\ref{two}) are classified by the cohomology $H^1({\cal H})$ of a bundle ${\cal H}$ that we will specify.  In Section \ref{Sec:sigma} we will see that, in the limit $\alpha' \rightarrow 0$, the equations above precisely duplicate the cocycle conditions found in the worldsheet analysis of \cite{ms}.

\section{Sequences, maps in cohomology and interpreting the fluctuations of the Strominger system}\label{Sec:seq}

We define a generalization of the Atiyah groupoid as follows. First we construct the Atiyah groupoid itself \cite{atiyah} associated to $V \oplus TX$. In other words, we will be interested in considering the holomorphy of the bundle which is the direct sum of the gauge and tangent bundles.
The pertinent extension is
\begin{align} \label{atiyah}
0 \to \textnormal{End}_0(V) \oplus \textnormal{End}_0(TX) \to  {\cal Q} \to TX \to 0,
\end{align} 
where, in the standard form for Atiyah structures, the extension class is
determined by
\begin{align}
F + \hat{R} \in H^1(\textnormal{End}_0(V) \otimes TX^{\vee}) \oplus
H^1(\textnormal{End}_0(TX) \otimes TX^{\vee}). \nonumber
\end{align}
We then define a further bundle, ${\cal H}$, in terms of ${\cal Q}$. This bundle (or rather its dual) has already been examined, in detail, in the context of Strominger systems by Baraglia and Hekmati, in a paper studying T-duality properties of the heterotic string \cite{Baraglia:2013wua}:
\begin{align} \label{groupoid}
0 \to TX^{\vee} \to {\cal H} \to {\cal Q} \to 0~.
\end{align}
In the limit $\alpha' \rightarrow 0$, the extension is determined by $\partial J$, and for nonzero $\alpha'$, by non-trivial combinations of $\partial J$, $F$ and $R$, as we shall discussion in Section \ref{good_seq_check}.
We claim the metric, spin connection, and bundle (``F-flat'') deformations of the
Strominger system, as described by the fluctuation analysis in the previous section, are given by $H^1({\cal H})$. The physical moduli are a subset of these fields determined by the ``D-term" equations (\ref{d1}) and a removal of the redundancy in the fluctuation of the spin connection.

In Subsection \ref{FTmatch} we will show that the above claim is true by comparing the cohomology group $H^1({\cal H})$ to the field theory analysis of the previous section. In doing so we will assume that there is a well defined extension (\ref{groupoid}) associated with the Strominger system. Once we have shown that the matching between unconstrained field fluctuations and cohomology groups described above holds, we will then return to the definition of (\ref{groupoid}) and discuss the nature of the extension class picked out by the supergravity data. We emphasize once more that the extension class for (\ref{groupoid}) associated to the Strominger system has been discussed in the literature in \cite{Baraglia:2013wua}.

\subsection{Matching to the field theory perturbation analysis} \label{FTmatch}

From the long exact sequence associated to (\ref{groupoid}) we find 
\begin{equation} \label{h1q}
H^0({\cal Q}) \to H^1(TX^{\vee}) \to H^1({\cal H}) \to H^1({\cal Q}) \to H^2(TX^{\vee})  \;.
\end{equation}
This leads to the following expression for the cohomology $H^1({\cal H})$:
\begin{equation} \label{cohfluct}
H^1({\cal H}) = \left\{ \begin{array}{c} \textnormal{ker} (H^1({\cal Q}) \to H^2(TX^{\vee})) \\ 
\oplus \\ 
\textnormal{coker} (H^0({\cal Q}) \to H^1(TX^{\vee}) ) \;. \end{array} \right.
\end{equation}
From the long exact sequence in cohomology associated to (\ref{atiyah}) we have that,
\begin{eqnarray} \label{h1Q}
H^1({\cal Q}) &= \left\{\begin{array}{c} \textnormal{coker}(H^0(TX) \to H^1(\textnormal{End}_0(V) )\oplus H^1(\textnormal{End}_0(TX)))\\ \oplus \\ \label{h0Q1} \textnormal{ker}( H^1(TX) \to H^2(\textnormal{End}_0(V))\oplus H^2(\textnormal{End}_0(TX)))\;, \end{array}\right. \\ \label{thirdone}
\textnormal{and} \;\; H^0({\cal Q}) &= \;\;\;\;\;\; \textnormal{ker}(H^0(TX) \to H^1(\textnormal{End}_0(V)) \oplus H^1(\textnormal{End}_0(TX))) \;.
\end{eqnarray}

Combining equations (\ref{cohfluct}), (\ref{h1Q}) and (\ref{thirdone}) we see that $H^1({\cal H})$ is a subspace  of $H^1(\textnormal{End}_0(V)) \oplus H^1(\textnormal{End}_0(TX)) \oplus H^1(TX) \oplus H^1(TX^{\vee}) $. One should think of these as the relevant fluctuations of the gauge connection, spin connection, complex structure, and Hermitian two form respectively. 

In order to see that the constraints in $(\ref{cohfluct})$ correspond to those seen in the field theory analysis of the proceeding section we will, in the next subsection, consider the case where $H^0(TX)=0$. This specialization is not necessary but simplifies the ensuing discussion, making the extraction of the salient points much easier. In subsection \ref{h0txnot0} we will return to the case with $H^0(TX)\neq 0$ to complete our discussion.

\subsubsection{The $H^0(TX)=0$ case} \label{h0tx0}

In cases where $H^0(TX)=0$ the cohomology $H^1({\cal H})$ simplifies as follows:
\begin{eqnarray} \label{cohfluctnew}
H^1({\cal H}) = \left\{ \begin{array}{c} \textnormal{ker} (H^1({\cal Q}) \to H^2(TX^{\vee})) \\ \oplus \\  H^1(TX^{\vee})\; . \end{array}\right.  
\end{eqnarray}
In this expression we have
\begin{eqnarray} \label{h1Qnew}
H^1({\cal Q}) &=& \left\{\begin{array}{c}  H^1(\textnormal{End}_0(V) ) \oplus H^1(\textnormal{End}_0(TX))\\ \oplus \\ \label{h0Q2} \textnormal{ker}( H^1(TX) \to (H^2(\textnormal{End}_0(V)) \oplus H^2(\textnormal{End}_0(TX))))\;.\end{array}\right. 
\end{eqnarray}

To compare these expressions with (\ref{one}), (\ref{two}) and (\ref{three}) we start by considering the expression for $H^1({\cal Q})$. 

First, note equation (\ref{two}) states that any allowed fluctuation in $\delta J \in H^1(TX)$ can be mapped by the field strength $F$ into a $\overline{D}$ exact form. Similarly, equation (\ref{three}) states that an allowed fluctuation maps via $R$ into an exact $\textnormal{End}_0(TX)$ valued two-form. This is precisely the content of the second line in (\ref{h1Qnew}), if we take the map involved to be given by the cohomology classes $[F]$ and $[\hat{R}]$. This is the standard Atiyah class story \cite{Anderson:2010mh,Anderson:2011ty,Anderson:2013qca,atiyah}, as applied to the direct sum of the holomorphic gauge and tangent bundles.

Next, let us write down the most general fluctuations in $A$ that are allowed by equation (\ref{two}). They take the form, 
\beq \label{split}
\delta A_{\overline{b}} = \delta A_{\overline{b}}^{\delta J} + \delta A_{\overline{b}}^0 \;,
\eeq
where $\delta A^{\delta J}$ is any specific chosen solution to (\ref{two}) and $\delta A^0$ is any $\overline{D}$ closed form which defines, up to gauge transformations, an element of $H^1(\textnormal{End}_0(V))$. Note that $\delta A^{\delta J}$ is only defined up to the addition of a closed piece, or, given the possibility of gauge transformations, up to the addition of an element of $H^1(\textnormal{End}_0(V))$. This will be of importance shortly. The contributions $\delta A^0$ are what are normally referred to as bundle moduli. By contrast, $\delta A^{\delta J}$ is a part of the reduction ansatz of the theory to four dimensions which describes how the gauge field adjusts to remain holomorphic under a change in complex structure of the base \cite{Anderson:2010mh,Anderson:2011ty,Anderson:2013qca}. The fluctuations $\delta A^0$ are, up to gauge transformations, exactly what is described by the first term in the first line of (\ref{h1Qnew}). An exactly analogous discussion can be made for the fluctuations of the spin connection, $\delta W$, in equation (\ref{three}).

\vspace{0.1cm}

The proceeding two paragraphs show that equations (\ref{two}) and (\ref{three}) describe precisely the content of (\ref{h1Qnew}). We shall now analyze the content of $H^1({\cal H})$ in equation (\ref{cohfluctnew}) and compare it to (\ref{one}).

\vspace{0.1cm}

We begin with the first line of (\ref{cohfluctnew}). 
This kernel contains two pieces, one lying inside 
\begin{equation}
H^1(\textnormal{End}_0 (V)) \oplus H^1(\textnormal{End}_0(TX)),
\end{equation}
and the other inside
\begin{equation}
H^1(TX) \;.
\end{equation} 
\begin{itemize}
\item We first consider the bundle moduli and spin connection fluctuation piece. Consider the solution to (\ref{two}) in (\ref{split}). We shall consider a variation where $\delta A^0$ is non-vanishing but where the perturbation of the complex structure (and so $\delta A^{\delta J}$) and spin connection are set to zero. Substituting this expression for the gauge field fluctuation into (\ref{one}) we find the following:
\beq \label{newbadger}
-\frac{4}{30}\alpha' \delta_{xy} \delta A^{0x}_{[\overline{a}}F^y_{\overline{b}]c} =  i \overline{\partial}_{[\overline{a}} \delta J_{\overline{b}]c}-2\overline{\partial}_{[\overline{a}} \delta B_{\overline{b}] c}  - \overline{\partial}_{[\overline{a}} \Lambda^{\alpha'}_{\overline{b}]c}~.
\eeq
The right hand side of (\ref{newbadger}) is of the form $\overline{\partial}_{[\overline{a}} \Gamma_{\overline{b}]c}$ for some $\Gamma$. Raising the index $c$ we then see that this is exactly the content of the bundle modulus dependent part of the first line of (\ref{cohfluctnew}), where we take the relevant map\footnote{One might be concerned that variations of the other two terms  on the left hand side of (\ref{one}) would allow for more general possible bundle modulus variations. This is indeed the case if one simultaneously considers other variations besides $\delta A^0$. This reflects the composite nature of the kernel present in (\ref{cohfluctnew}) and is also reproduced by the sequence structure. We are treating the various types of perturbation independently here for ease of exposition.} to be given by $-\frac{4}{30} \alpha'[F]$. Exactly parallel comments can be made for the fluctuation of the spin connection and the third term in equation (\ref{one}).
\item The piece of the first line of (\ref{cohfluctnew}) lying inside $H^1(TX)$ can be understood in a similar, albeit less direct, fashion. At zeroth order in $\alpha'$ the first term on the left of (\ref{one}) is the one which depends directly on complex structure variation. We then see that this must be equal to an exact $(1,1)$ form. Raising the index $c$ with the metric this is precisely the content of the $H^1(TX)$ contribution to $H^1(Q)$ in the kernel in (\ref{cohfluctnew}) where the relevant map is taken to be $\partial J$. (As a map in cohomology we can equally take this zeroth order map to be $2 iH = (\partial - \overline{\partial})J$ as the $\overline{\partial} J$ term is trivial in cohomology.)

At first order in $\alpha'$ this discussion gets somewhat modified. Via the solution (\ref{split}) to equation (\ref{two}) the second term in (\ref{one}) also depends on the complex structure variation (the reduction ansatz for the gauge field in heterotic compactifications depends upon complex structure moduli as has been described in \cite{Anderson:2010mh,Anderson:2011ty,Anderson:2013qca}). Similarly, the third term in equation (\ref{one}) also depends on $\delta J$ via equation (\ref{three}). In fact, keeping $\delta A^0$ and the analogous quantity for the spin connection $\delta W^0$ fixed, we find the following:
\beq \label{onemark2}
\delta J_{[\overline{a}}^{\;d}  \partial J_{\overline{b}]cd} -\frac{4}{30} \alpha' \delta_{xy} \delta A^{x\delta J}_{[\overline{a}} F^y_{\overline{b}]c} + 4 \alpha' \delta W^{\delta J \;\alpha \beta}_{[\overline{a}} R^{\beta \alpha}_{\overline{b}]c} ~~=~~  \textnormal{``exact."}
\eeq
This expression should be regarded as a complicated linear map of the complex structure fluctuation $\delta J \in H^1(TX)$ under consideration - with the complex structure dependence of $\delta A^{\delta J}$ and $\delta W^{\delta J}$ being defined by (\ref{two}) and (\ref{three}) respectively. Comparison to the right hand side of (\ref{onemark2}) tells us that $\delta J$ should be taken by this implicitly defined map to an exact form. This defines the order $\alpha'$ map between $H^1(TX)$ (inside $H^1({\cal Q})$) and $H^2(TX^{\vee})$ in (\ref{cohfluctnew}) to first order in $\alpha'$. We define a form $M$ by 
\beq \label{Mdef}
\delta J_{[\overline{a}}^{\;d}  \partial J_{\overline{b}]cd} -\frac{4}{30} \alpha' \delta_{xy} \delta A^{x\delta J}_{[\overline{a}} F^y_{\overline{b}]c} + 4 \alpha' \delta W^{\delta J \;\alpha \beta}_{[\overline{a}} R^{\beta \alpha}_{\overline{b}]c} ~=~ \delta J_{[\overline{a}}^{\;d} M_{\overline{b}]cd}
\eeq
so that (\ref{onemark2}) becomes
\beq  \label{onemark3}
\delta J_{[\overline{a}}^{\;d} M_{\overline{b}]cd} = \textnormal{``exact."}
\eeq
Note that there must be a free index on $\delta J$ on the left hand side of (\ref{onemark3}) as there is in (\ref{one}), (\ref{two}) and (\ref{three}). Were this not to be the case then, for specific choices of the values of the free indices, the right hand side of (\ref{Mdef}) would depend upon components of $\delta J$ which do not appear on the left hand side. That this is indeed a good map in cohomology to this order will be shown in the next subsection, where we will also be somewhat more explicit about the nature of $M$.
\end{itemize}

All that remains for us to do in showing that the content of (\ref{one}), (\ref{two}) and (\ref{three}) is the same as that of (\ref{cohfluctnew}) and (\ref{h1Qnew}) is to demonstrate that the allowed Hermitian fluctuations ($(1,1)$ perturbations of the two form $J$, together with any associated compensating variations in the reduction ansatz for the other fields) are indeed counted by $H^1(TX^{\vee})$. This is almost clear from (\ref{one}), however the third term on the left hand side would require some algebra to analyze properly in this regard. For this piece of the analysis it is, in fact, easier to recombine the equations we split up with the introduction of $H$ at the start of the paper and simply consider such fluctuations in equation (\ref{bi1}) (which has the same content as (\ref{one})). As has already been noted in \cite{Gillard:2003jh}, allowed $(1,1)$ variations of $J$ in this equation correspond to Aeppli cohomology classes \cite{angella}. This is simply because variations in the right hand side are $d$ exact and can, therefore, by the $\partial \overline{\partial}$-lemma be written as $\partial \overline{\partial}$ of something. The allowed fluctuations (keeping complex structure fixed as we have already discussed its variation) are therefore a combined fluctuation of $J_{(1,1)}$ and the gauge field (the latter introducing a new piece into the reduction ansatz of the theory as (\ref{two}) did for variations in the complex structure) which is $\partial \overline{\partial}$ closed. Modding out by changes which can be induced by coordinate transformations this leads to the fluctuations being counted by the Aeppli cohomology group. However, on a $\partial \overline{\partial}$-manifold the Aeppli and Dobeault cohomology groups are isomorphic \cite{deligne,aeppli}, hence the allowed fluctuations in the Hermitian moduli are exactly as described in the second line of (\ref{one}).

In short, the allowed fluctuations of the Hermitian two form, gauge field, NS two form, complex structure and spin connection, under variation of the ``F-term" equations in the Strominger system, (\ref{one}) and (\ref{two}),
are exactly characterized by $H^1({\cal H})$ as defined in (\ref{groupoid}). In the next subsections, we will show that all of the maps we have derived are good maps in cohomology and we will complete our discussion in the case where $H^0(TX)\neq 0$. In addition, we will show that the extension class associated to (\ref{groupoid}), defined by the Strominger system, is indeed an element of the correct cohomology group.

\subsubsection{Well-definedness of the map in cohomology}\label{well_defined}

We have established that the tangent to the moduli space of the Strominger system is a subspace of $H^1({\cal H})$ as given by (\ref{cohfluctnew}) and (\ref{h1Qnew}). We must now demonstrate that the maps in these expressions, as given by (\ref{one}), (\ref{two}) and (\ref{three}), are good maps in cohomology. In the case of the Atiyah groupoid, (\ref{h1Qnew}) and (\ref{two}) and (\ref{three}), this is already well known and established \cite{Anderson:2010mh,Anderson:2011ty,Anderson:2013qca,atiyah}. Thus, we need only focus on the map in (\ref{cohfluctnew}) and (\ref{one}).

For a map to be well defined between cohomologies the following properties should hold:
\begin{enumerate}
\item the image does not depend upon the representative used to describe the element of the source,
\item the image of a closed form is a closed form,
\item the map on cohomology is gauge invariant.
\end{enumerate}

\subsubsection*{Zeroth order in $\alpha'$}\label{james_order_zero}

To zeroth order in $\alpha'$ the structure is easy to verify. We reiterate the structure we are investigating at this order here for ease of presentation.
\begin{eqnarray} \label{zeroth}
H^1({\cal H}) = \left\{ \begin{array}{c} \textnormal{ker} \left(\textnormal{ker}\{H^1(TX) \stackrel{[F], [R]}{\longrightarrow} H^2(\textnormal{End}_0(V)) \oplus H^2(\textnormal{End}_0(TX))\} \stackrel{\partial J}{\longrightarrow} H^2(TX^{\vee})\right) \\ \oplus \\ H^1(\textnormal{End}_0(V)) \oplus H^1(\textnormal{End}_0(TX)) \oplus H^1(TX^{\vee}) \;. \end{array} \right.  
\end{eqnarray}
Note that the non-trivial maps acting on the bundle moduli and spin connection fluctuations are order $\alpha'$ in (\ref{one}) and thus drop out above.

At this order in $\alpha'$ the Bianchi identity is simply $\partial \overline{\partial} J=0$. From here it is trivial to see that a form $\delta J_{\overline{a}}^c = \overline{\nabla}_{\overline{a}}v^c$ for some $v^c$ maps to an exact form, and thus the map image does not depend upon the representative used in a given class. The same Bianchi identity also makes it clear that the map $\partial J$ always takes the source to closed forms. Finally the map is clearly gauge invariant under all symmetries in the problem and thus, at zeroth order in $\alpha'$, the maps we have obtained are well defined between the cohomology groups.

\subsubsection*{First order in $\alpha'$}
At first order in $\alpha'$ the map $\partial J$ is replaced by $M$, as
implicitly encoded in equations (\ref{onemark2}) and (\ref{onemark3}).  In addition the maps on $\delta A^0$ in (\ref{split}) appearing in (\ref{one}), and the analogous structure for the perturbations in $H^1(\textnormal{End}_0(TX))$, are non-zero at this order in $\alpha'$. Given all of this, the structure we now have is as follows.
\begin{eqnarray} \label{order1}
H^1({\cal H}) = \left\{ \begin{array}{c} \textnormal{ker} \left(\textnormal{ker}\{H^1(TX) \stackrel{[F],[R]}{\longrightarrow} H^2(\textnormal{End}_0(V)) \oplus H^2(\textnormal{End}_0(TX))\} \stackrel{M}{\longrightarrow} H^2(TX^{\vee})\right) \\ \oplus \\ \textnormal{ker}\left( H^1(\textnormal{End}_0(V)) \stackrel{-\frac{4}{30} \alpha' [F]}{\longrightarrow} H^2(TX^{\vee}) \right) \oplus \textnormal{ker}\left( H^1(\textnormal{End}_0(TX)) \stackrel{4\alpha' [R]}{\longrightarrow} H^2(TX^{\vee}) \right)~~ \\ \oplus \\ H^1(TX^{\vee}) \;. \end{array}\right.
\end{eqnarray}

To the order in $\alpha'$ at which we are working, the maps $-\frac{4}{30}\alpha' [F]$ and $4\alpha'[R]$ can trivially be shown to be both maps into closed forms and independent of representatives of the source element used within a class in $H^1(\textnormal{End}_0(V))$ or $H^1(\textnormal{End}_0(TX))$ by a simple use of the Bianchi identities $DF=0$ and $\hat{\nabla} \hat{R}=0$. Thus the second line in (\ref{order1}) is well defined.

The Atiyah map, determined by $[F]$ and $[\hat{R}]$, in the first line of (\ref{order1}) is essentially unchanged from zeroth order and is well known to be well defined. Thus we need only analyze the map $M$, as given in (\ref{Mdef}).

To see that $M$ always maps to closed forms we simply take the $\overline{\nabla}$ exterior derivative of the left hand side of (\ref{Mdef}). We find,
\begin{eqnarray} \label{Mclosed1}
&&\delta J_{[\overline{a}}^{\;d} \overline{\nabla}_{\overline{c}} \partial J_{\overline{b}]cd} - \frac{4}{30} \alpha' \delta_{xy} \overline{D}_{[\overline{c}} \delta A^{x\delta J}_{\overline{a}} F^y_{\overline{b}]c} + 4 \alpha' \overline{\nabla}_{[\overline{c}} \delta W^{\delta J \, \alpha \beta}_{\overline{a}} R^{\beta \alpha}_{\overline{b}]c} \\ 
\nonumber &=& \delta J_{[\overline{a}}^{\;d} \overline{\nabla}_{\overline{c}} \partial J_{\overline{b}]cd} - \frac{2}{30}i \alpha' \delta_{xy} \delta J_{[\overline{c}}^{\;d} F^x_{\overline{a} |d| } F^y_{\overline{b}]c} + 2 \alpha' i \delta J_{[\overline{c}}^{\;d} R_{\overline{a}|d|}^{\alpha \beta} R^{\beta \alpha}_{\overline{b}]c} \;, \\ \nonumber
&=& \frac{1}{2}  \delta J_{[\overline{a}}^{\;d} \left( \frac{1}{30} i \alpha' \textnormal{tr} F \wedge F - i \alpha' \textnormal{tr} R \wedge R \right)_{\overline{b} \overline{c}] cd} -\alpha' \frac{1}{2} \frac{1}{30} i\delta J_{[\overline{c}}^{\;d} ( \textnormal{tr} F\wedge F)_{\overline{a} \overline{b}]cd} + i  \frac{1}{2} \alpha' \delta J_{[\overline{c}}^{\;d} (\textnormal{tr} R \wedge R)_{\overline{a} \overline{b}]cd} \;, \\ \nonumber
&=& 0 \;.
\end{eqnarray}
Thus we find that the left hand side of (\ref{Mdef}) is closed as desired. In (\ref{Mclosed1}) we have used the Bianchi identity for the gauge field, the heterotic Bianchi identity, and equations (\ref{two}) and (\ref{three}). 

It should be noted that the calculation in (\ref{Mclosed1}), together with the proceeding discussion in this section, shows that we can write $M$ as
\beq\label{mdef}
M = \partial J + i \frac{1}{30} \alpha' \omega_3^{\textnormal{YM}} - i \alpha' \omega_3^{\textnormal{L}} + M^0 \;,
\eeq
where $M^0$ is a $\overline{\partial}$-closed from. This form of $M$ will be used in the next section to make connections with the work of \cite{Baraglia:2013wua,garcia_ferndandez}.

That the map $M$ is independent of the representative of the class in $H^1(TX)$ of $\delta J$ can be shown using the gauge field, spin connection and heterotic Bianchi identities and is left as an exercise for the reader. Gauge invariance of $M$ as a map in cohomology is just a straightforward to demonstrate, making use of the $\partial \overline{\partial}$-lemma.

\subsubsection{Including non-vanishing $H^0(TX)$ and coordinate transformations} \label{h0txnot0}

Consider the infinitesimal coordinate transformation
\beq\label{coordtrans}
z^a= z'^a + v^a(z',\overline{z}') \;.
\eeq
Under such a transformation the $(0,2)$ part of the field strength has the following first order variation in $v^a$
\begin{align} \label{fchange}
 \delta_v F_{\overline{a}' \overline{b}'} &=~ \frac{\partial F_{\overline{a}' \overline{b}'}}{\partial z^c} v^c + \frac{\partial F_{\overline{a}'\overline{b}'}}{\partial \overline{z}^{\overline{c}}} \overline{v}^{\overline{c}} +  \frac{\partial v^c}{\partial \overline{z}^{\overline{a}'}} F_{c \overline{b}'} + \frac{\partial v^c}{\partial \overline{z}^{\overline{b}'}} F_{\overline{a}'c} =0 \;, \\ 
&=~   \frac{\partial v^c}{\partial \overline{z}^{\overline{a}'}} F_{c \overline{b}'} + \frac{\partial v^c}{\partial \overline{z}^{\overline{b}'}} F_{\overline{a}'c} \;, \\
&=~ 2 \partial_{[\overline{a}'} \left( v^a F_{|a| \overline{b}']}\right) - 2 v^a \partial_{[\overline{a}'} F_{|a| \overline{b}']} \;, \\
&=~ 2 \partial_{[\overline{a}'} \left( v^a F_{|a| \overline{b}']}\right) - 3 v^a \partial_{[\overline{a}'} F_{a \overline{b}']} \;.
\end{align}
Now we use the Bianchi identity on the Yang-Mills field strength, $DF=0$, to simplify the second term:
\begin{align}
 \delta_v F_{\overline{a}' \overline{b}'}  &=~ 2 \partial_{[\overline{a}'} \left( v^a F^x_{|a| \overline{b}']}\right) - 3 v^a \left(- f^x_{yz} A^y_{[\overline{a}'} F^z_{a \overline{b}']} \right) \;, \\ 
&=~ 2 \partial_{[\overline{a}'} \left( v^a F^x_{|a| \overline{b}']}\right) +2 v^a \left(f^x_{yz} A^y_{[\overline{a}'} F^z_{|a| \overline{b}']} \right)\;,  \\ \label{lastone}
&=~ 2 D_{[\overline{a}'} \left(v^a F_{|a| \overline{b}']} \right) \;.
\end{align}

If $v \in H^0(TX)$ then, using the gauge field Bianchi identity, equation (\ref{lastone}) is equal to zero. In such a situation, $v^a F_{a \overline{b}'}$ defines a perturbation to the gauge field which is an element of $H^1(\textnormal{End}_0(V))$ and which can be obtained by a simple coordinate transformation of the field strength. Such an element should not be considered as a separate bundle modulus degree of freedom. This explains the presence of the $\textnormal{coker}$ in equation (\ref{h1Q}) in situations where $H^0(TX)\neq0$.

If $v^a F_{a \overline{b}'}$ is exact, then this degree of freedom $v^a$ clearly does not remove a bundle modulus degree of freedom (it would map to a zero element in the relevant cohomology). The set of such vectors, which also can not be used to remove a spin connection degree of freedom, is counted by $H^0({\cal Q})$ as described in equation (\ref{thirdone}). Such unaccounted for $v^a$ can be used, via a coordinate transformation of the form (\ref{coordtrans}), to remove a supposed Hermitian modulus. This explains the coker in the second line of equation (\ref{cohfluct}). 

Thus all of the appearances of $H^0(TX)$ in $H^1({\cal H})$ simply account for the removal of some moduli degrees of freedom as simple coordinate transformations. We have therefore proved that the cohomology $H^1({\cal H})$ does indeed count the ``F-flat" perturbations in the Hermitian two form, complex structure, gauge connection, spin connection and NS two form of the Strominger system.

\subsubsection{Well definedness of the bundle ${\cal H}$}\label{good_seq_check}

If the formalism we have presented is to make sense, we must demonstrate that the bundle ${\cal H}$ can be well defined in terms of the supergravity data. The content of this subsection has already been discussed in the work of Baraglia and Hekmati \cite{Baraglia:2013wua}.

The extension (\ref{groupoid}) is controlled by the extension group $\textnormal{Ext}^1({\cal Q},TX^{\vee})=H^1({\cal Q}^{\vee}\otimes TX^{\vee})$. Taking the dual of the sequence (\ref{atiyah}), and tensoring it by $TX^{\vee}$ we have the following:
\beq
0\to TX^{\vee} \otimes TX^{\vee} \to {\cal Q}^{\vee} \otimes TX^{\vee} \to (\textnormal{End}_0(V) \oplus \textnormal{End}_0(TX))\otimes TX^{\vee} \to 0 \;.
\eeq
Examining the associated long exact sequence in cohomology we discover that $H^1({\cal Q}^{\vee}\otimes TX^{\vee})$ receives contributions from two terms.
\begin{eqnarray} 
\lefteqn{
H^1({\cal Q}^{\vee} \otimes TX^{\vee})  
}
\nonumber \\
& = & \left\{ \begin{array}{c} \textnormal{ker}( (H^1(\textnormal{End}_0(V) \otimes TX^{\vee}) \oplus H^1(\textnormal{End}_0(TX) \otimes TX^{\vee})) \to H^2(TX^{\vee} \otimes TX^{\vee})) \\ \oplus \\
\textnormal{coker}( (H^0(\textnormal{End}_0(V) \otimes TX^{\vee}) \oplus H^0(\textnormal{End}_0(TX) \otimes TX^{\vee})) \to H^1(TX^{\vee} \otimes TX^{\vee}))~
\;.~~
\end{array} \right. \label{thischap}
\end{eqnarray}
As with the maps in cohomology discussed in the proceeding subsections, the extension defined by the Strominger system is determined by the map that appears in equation (\ref{one}). 
More precisely, the part of the extension in the kernel of the first line of (\ref{thischap}) is determined by 
\begin{equation}
-\frac{4}{30}\alpha' [F] 
\in H^1(\textnormal{End}_0(V) \otimes TX^{\vee}), \; \; \;
4 \alpha' [\hat{R}] \in H^1(\textnormal{End}_0(TX) \otimes TX^{\vee}). 
\end{equation}
To disambiguate, note that
the map itself is also a combination of $[F]$ and $[\hat{R}]$, albeit
without additional factors.  We then find that the image of the first map in (\ref{thischap}) is $-\frac{4}{30} \alpha' {\rm tr} F\wedge F + 4\alpha' {\rm tr}R \wedge R$ which is indeed zero in $H^1(\textnormal{End}_0(V) \otimes TX^{\vee})$ by the Bianchi identity (\ref{bi1}). 

The contribution to the extension class defined by the Strominger system which lies in the cokernel in the second line on (\ref{thischap}) is somewhat more complicated to extract from (\ref{one}), (\ref{two}) and (\ref{three}). One might think that the natural mapping appearing in this equation, with index structure compatible with being an element of $H^1(TX^{\vee} \otimes TX^{\vee})$, is simply $\partial J$. This, however is not a $\overline{\partial}$ closed form and as such is not in the cokernel piece of (\ref{thischap}). As with the map in cohomology defined in (\ref{onemark2}), the key observation is that the first term on the left hand side of (\ref{one}) is not the only term to depend upon $\delta J$. Due to the relations (\ref{two}) and (\ref{two}) the other two terms do as well.

In Section \ref{h0tx0} we defined the map $M$ implicitly using the expression in (\ref{Mdef}). This $M$ is, up to raising and lowering indices, an element of $H^1(TX^{\vee} \otimes TX^{\vee})$ and is, in fact, the object which defines the portion of the extension class lying in the cokernel in the second line of (\ref{thischap}). To see why the cokernel structure comes about from a field theory perspective, we need to consider the ambiguity in the definition of $\delta A^{\delta J}$ discussed in the text underneath (\ref{split}) and the relationship between $\delta A^{\delta J}$ and $M$ given in (\ref{Mdef}). The freedom of the definition of $\delta A^{\delta J}$ includes the ability to add a closed piece of the form $\delta J_{\overline{b}}^a v^x_a$, where $v^x_a$ is a closed endomorphism valued $(1,0)$ form. By (\ref{onemark2}) this changes $\delta J M$ by a piece $-\frac{4}{30} \alpha' \delta_{xy} \delta J_{[\overline{a}}^a v^x_{|a|}
 F^y_{\overline{b}]c}$. Such a change would lie precisely in the image of the map in the second line of (\ref{thischap}). Given this, and the proceeding discussion in this subsection, the extension class given by the Strominger system does indeed form an element of $H^1({\cal Q}^{\vee} \otimes TX^{\vee})$.

It is important to note that the appearance of $\textnormal{End}_0(TX)$ in the definition of ${\cal Q}$ is vital for the extension ${\cal H}$ to be well defined. This is the equivalent in the sequence based discussion of treating the spin connection, Hermitian two form and complex structure deformations separately in the field theory analysis.

\subsection{Observations on the structure presented in this section}

In this subsection we will discuss several immediate consequences of the structure described above. As we will show in Section \ref{Sec:sigma}, the results we have presented above reduce to those of \cite{ms}, which were derived formally from two-dimensional non-linear sigma models, in the limit where $\alpha' \to 0$. In such a limit, as we have seen, the maps involved in defining ${\cal H}$ become significantly more straightforward and, thus, it is interesting to look at the new features arise when we go beyond such a simplification. 

One such new feature is that, unlike in the Atiyah class discussion of \cite{Anderson:2010mh,Anderson:2011ty,Anderson:2013qca,atiyah}, the bundle moduli of Strominger systems are constrained by the map structure in (\ref{order1}). This is, in fact crucial to note in determining the correct sequence structure to reproduce the fluctuation analysis culminating in equation (\ref{one}). Many promising candidates, for example,
\beq
0 \to TX^{\vee} \oplus \textnormal{End}_0(V) \to \tilde{{\cal H}} \to TX \to 0 \;,
\eeq
are not compatible with the supergravity analysis due to a failure to reproduce this restriction.

The constraint which we obtain in (\ref{order1}) on the bundle moduli is that they must be in the kernel of a map defined by $-\frac{4}{30}\alpha' [F]$. Presumably, this constraint is reproduced in the four dimensional effective theory by the requirements $\partial W=0$ where the relevant piece of the heterotic Gukov-Vafa-Witten superpotential is $\int \Omega \wedge \omega^{\textnormal{YM}}_3$. This complements the appearance of the Kuranishi obstruction to complex structure moduli in the theory, 
which is well known to appear via the equation $W=0$ for the same superpotential \cite{Berglund:1995yu}[section 2.3].

If one regards the low energy fields as elements of $H^1(TX) \oplus H^1(TX^{\vee}) \oplus H^1(\textnormal{End}_0(V))\oplus H^1(\textnormal{End}_0(V))$, the upper bound on the number of these degrees of freedom which can be stabilized is determined by the dimension of the targets of the maps in equation (\ref{order1}). For stabilization by a holomorphic bundle on a Calabi-Yau this has been emphasized in \cite{Anderson:2010mh,Anderson:2011ty,Anderson:2013qca}, and leads to a maximum of $h^1(\textnormal{End}_0(V))$ complex structure moduli being stabilized. An examination of (\ref{order1}) reveals that we can do substantially better than this in the Strominger system.

Tallying the dimensions of the various target spaces, we see that in principle all of the complex structure and bundle moduli can be stabilized in the Strominger system. The Hermitian moduli are, however, unrestricted by these ``F-term" constraints. 

We note that we have only considered equations (\ref{mrbi}), (\ref{first1}) and (\ref{mrhol})  and not the equations (\ref{d1}) in discussing the allowed fluctuations of the Strominger system on a manifold obeying the $\partial \overline{\partial}$-lemma. Thus the actual massless degrees of freedom of the four dimensional theory are a subset of those we are discussing here. Thus one can not conclude from our results that there are $h^1(TX^{\vee})$ massless Hermitian moduli in such a compactification. Rather, one only knows that the massless Hermitian moduli are a subset of those classified by this finite cohomology group.

It should be noted that the overall scale of the compactification does not correspond to one of these moduli, except in the very simplest of cases. Rescaling the compactification space corresponds to an operation of the form $J \to (1+\epsilon) J$ such that we have $\delta J= \epsilon J$ for some small $\epsilon$. Such a change $\delta J$ is not an element of $H^1(TX^{\vee})$ except in the special case where $J$ is $\overline{\partial}$-closed. $J$ is a real form and, as such, this also corresponds to $J$ also being $\partial$ closed. Thus, by equation (\ref{first}), the overall volume of the compactification is only a modulus if $H=0$.

Given the similarities in the structure we see here to Hermitian Yang-Mills and its relation to the Atiyah groupoid, it is natural to conjecture that the equations we have analyzed, (\ref{mrbi}-\ref{mrhol}), correspond to F-term restrictions in the four dimensional theory, were as the equations (\ref{d1}) correspond to D-terms. If this is indeed the case, then the constraints on the remaining degrees of freedom can be determined purely from their charges under the four dimensional gauge group (in addition to the associated K\"ahler potential). The proof that equations (\ref{d1}) do indeed correspond to D-term constraints in the four dimensional effective theory will be attempted in future work \cite{usfuture}.

As a final comment we will note that, although we have only described the analogue of what are usually thought of as uncharged moduli here, our discussion equally well applies to describing the massless matter content of a Strominger compactifications on a $\partial \overline{\partial}$-manifold. Such fields can simply be included by taking $V$ to be an $E_8$ bundle with reduced structure group. A subset of the moduli of this object are what are normally regarded as matter fields (i.e. matter degrees of freedom simply correspond to rank changing deformations of the what is usually regarded as the vector bundle of the system).

\section{Links to algebroids and Hitchin's generalized geometry} \label{Sec:algebroids}

In this section we explore the link between the fluctuation analysis in \eref{one}-\eref{three}, the short exact sequences in Section \ref{Sec:seq} and the structure of transitive Courant algebroids.

\subsection*{Transitive Courant algebroids}
The short exact sequences of bundles in \eref{atiyah} and \eref{groupoid} are closely related to a mathematical structure known as a ``transitive Courant algebroid". Courant algebroids \cite{bressler1,liu_xu,vaisman,severa} are familiar in the context of generalized geometry 
(as introduced by Hitchin \cite{first_hitchin}, see also {\it e.g.} \cite{cavalcanti_review,Koerber:2010bx}) and more recently have had relevance in the theory of reduction \cite{bcg} and in exceptional generalized geometry \cite{old_baraglia,rubio}. Much as the Atiyah algebroid of \eref{atiyah} encodes the simultaneous deformations of a bundle and its base manifold, intuitively, a Courant algebroid can be thought of as a mechanism for encoding deformations of linked structures. We will see in more detail in this section and Section \ref{Sec:sigma} the types of structures and coupled deformation problems that can be described by Courant algebroids.

A Courant algebroid is defined by the following: a vector bundle $V \to X$ over a smooth manifold $X$, a bilinear operator $[,]: \Gamma(V) \otimes \Gamma(V) \to \Gamma(V)$ on the space of sections of $V$, a non-degenerate bilinear form, $\langle , \rangle$ on $V$ and a bundle map $\rho: V \to TX$, called the ``anchor map." The data $(V, \rho,[,], \langle , \rangle )$  is called a ``Courant algebroid" if the following conditions\footnote{Here the differential in $d \langle a, b \rangle$ is built from $\kappa^{-1} \rho^{*} d$ where $\kappa: V \to V^{\vee}$ is induced by the inner product, $\rho^*$ is the dual to the anchor map and $d$ is the de Rham differential.} hold for all $a,b,c \in \Gamma(V)$ (see Appendix \ref{Courant_app} for further details):
\begin{itemize} 
\item $[a, [b,c]]=[[a,b],c]+[b,[a,c]]$,
\item $[a,b] + [b,a]=2d \langle a, b \rangle$,
\item $\rho(a) \langle b, c \rangle = \langle [a,b],c \rangle + \langle b, [a,c] \rangle$.
\end{itemize}
(see \cite{Hull:2009zb} for more on the physics of Courant brackets).

A Courant algebroid is transitive if the anchor $\rho$ is surjective, giving rise to an exact sequence of vector bundles
\beq
0 \to {\cal K} \to V \stackrel{\rho}{\longrightarrow} TX \to 0 \;,
\eeq
where ${\cal K}=ker(\rho)$. A very similar structure is already familiar to us through the Atiyah sequence \eref{atiyah} discussed in Section \ref{Sec:seq} and studied in our previous work \cite{Anderson:2010mh,Anderson:2011ty,Anderson:2013qca}. Indeed, given a principal bundle ${\cal P}$, it is possible to define the short exact sequence
\beq
0 \to {\mathfrak g}_{{\cal P}} \to {\cal A} \to TX \to 0 \;,
\eeq
where ${\mathfrak g}_{\cal P}$ is the adjoint bundle associated to ${\cal P}$ (in the case of $SU(n)$ bundles ${\mathfrak g}_{\cal P}=End_{0}(V)$ where $V$ is the rank $n$ bundle in the fundamental). 

The sequences defined in \eref{atiyah}-\eref{groupoid} are closely related to Courant algebroids and  have already been applied to heterotic theories in the context of generalized geometry. In this context, \cite{Baraglia:2013wua, garcia_ferndandez} defined a {\it heterotic Courant algebroid} to be a principal $G$-bundle ${\cal P}$ such that ${\cal K}={\cal U}/TX^{\vee}$ is isomorphic to the Atiyah algebroid of ${\cal P}$ (as a quadratic Lie algebroid) for some principal bundle ${\cal P}$. This definition naturally leads to the following sequences,
\begin{align}
& 0 \to {\cal K} \to {\cal U} \to TX \to 0 \;, \label{courant_alg1} \\
& 0 \to TX^{\vee} \to {\cal K} \to {\mathfrak g}_{\cal P} \to 0 \;. \label{courant_alg2}
\end{align}
It can be noted immediately that this has the same form as the dual of the sequences defined in \eref{atiyah}-\eref{groupoid}. We will return to this comparison momentarily, but first we must explore a bit more structure. A ``splitting" of ${\cal U}$ is a section $s: TX \to {\cal U}$ of the anchor $\rho$ such that the image $s(TX) \subset {\cal U}$ is isotropic\footnote{Recall that a quadratic form is said to be isotropic if there is a non-zero vector on which the form evaluates to zero. A subspace is isotropic if it contains some isotropic vector.} with respect to the pairing $\langle , \rangle$ on ${\cal U}$. 

Given a principal bundle ${\cal P}$ it is not always possible to define \eref{courant_alg1}, in general there is an obstruction for a quadratic Lie algebroid ${\cal A}$ to arise from a transitive Courant algebroid ${\cal U}$ as a quotient ${\cal U}^{\vee}/TX^{\vee}$. In fact, as has been shown in \cite{bressler1}, ${\cal A}$ comes from a transitive Courant algebroid ${\cal U}$ if and only if the first Pontryagin class vanishes, $p_1({\cal U})=0$ (see \cite{bouwknegt} for a review). It is important to note here that the first Pontryagin class is not in general equal to the second Chern class, but rather can be defined for any choice of pairing $\langle , \rangle$ (which crucially, may include non-trivial choices of sign/constant coefficients, see Appendix \ref{Courant_app}).

To understand this intuitively, consider that the non-trivial extension sequences of the form \eref{courant_alg1} are parameterized (up to isomorphism) by extension classes of the form
\beq
\textnormal{Ext}^{1}(TX, {\cal K})=H^1(X, TX^{\vee} \otimes {\cal K})
\eeq
(since $TX$ and ${\cal K}$ are smooth vector bundles). As in Section \ref{good_seq_check}, we can consider whether this extension class is non-trivial and hence whether or not it is possible to define a non-split sequence \eref{courant_alg1}. To evaluate this cohomology group we must consider the defining sequence of ${\cal K}$, twisted by $TX^{\vee}$:
\beq\label{twist_seq}
0 \to TX^{\vee} \otimes TX^{\vee} \to TX^{\vee} \otimes {\cal K} \to TX^{\vee} \otimes {\mathfrak g}_{{\cal P}} \to 0 \;.
\eeq
Taking the long exact sequence in cohomology leads to the following form for $H^1(X, TX^{\vee} \otimes {\cal K})$
\begin{align}
H^1(X, TX^{\vee} \otimes {\cal K})=& \textnormal{coker}(H^0(X, TX^{\vee} \otimes {\mathfrak g}_{{\cal P}}) \to H^1(X,TX^{\vee} \otimes TX^{\vee}))\label{first_half} \\
& \oplus \textnormal{ker}(H^1( X, TX^{\vee} \otimes {\mathfrak g}_{{\cal P}}) \to H^2(X, TX^{\vee} \otimes TX^{\vee}))\;.\label{second_half}
\end{align}
Focusing first on the kernel contribution above, it is clear that $[F^{1,1}] \subset H^1(TX^{\vee} \otimes {\mathfrak g}_{{\cal P}})$ that is, the space $H^1(TX^{\vee} \otimes {\mathfrak g}_{{\cal P}})$ is the space containing possible field strengths of the holomorphic vector bundle $V$. The co-boundary map and extension class defining \eref{twist_seq}-\eref{second_half} is simply the background field strength $[F_{0}^{1,1}]$. Hence, the condition that an element of $H^1(X,TX^{\vee} \otimes {\mathfrak g}_{{\cal P}})$ is in the kernel defined in \eref{second_half} is simply that (up to constant factors)
\beq\label{almost_anom}
Tr(F \wedge F) \sim {\bar \partial} H
\eeq
for some $H \in H^1(TX^{\vee} \otimes TX^{\vee})$. That is, that its first Pontryagin class vanishes.

The form of \eref{almost_anom} is suggestively close to the heterotic anomaly cancellation condition \eref{bi1}. In fact, it was this similarity that first led to the possible application of transitive Courant algebroids to heterotic theories \cite{Baraglia:2013wua, garcia_ferndandez}.

To make this explicit and to relate it to the study of fluctuations and moduli undertaken in this work, we will briefly review the arguments of \cite{Baraglia:2013wua, garcia_ferndandez} here. Recalling that the anomaly cancellation condition is not \eref{almost_anom} but rather ${\bar \partial }\partial J \sim {\rm tr}(R \wedge R)-{\rm tr}(F\wedge F)$, it is clear that the underlying $E_8 \times E_8$ bundle in a heterotic theory cannot play the role of the principal bundle ${\cal P}$ in \eref{courant_alg1}-\eref{courant_alg2} since ${\rm tr}(F\wedge F)$ is necessarily non-vanishing for this bundle. However, as shown in \cite{Baraglia:2013wua, garcia_ferndandez}, the structure of Courant algebroids can be applied by considering not the principal heterotic gauge bundle, but rather the direct sum of the gauge bundle and the principal frame bundle of the manifold $X$:
\beq\label{vtot}
{\cal V}_{total}={\cal V}_{gauge} \oplus {\cal V}_{frame}
\eeq
with structure group $G \times SO(6)$. For the case of $G=SU(n)$ we can define the Atiyah algebroid 
\beq
0 \to \textnormal{End}_0(V) \oplus \textnormal{End}_0(TX) \to {\cal A} \to TX \to 0 \;.
\eeq
Now to define the transitive Courant algebroid
\beq\label{the_goal}
0 \to {\cal A}^{\vee} \to {\cal U} \to TX \to 0 \;,
\eeq
the obstruction must vanish: $p_1({\cal U})=0$. But here following the same arguments that lead to \eref{almost_anom} lead to the condition
\beq
(a) {\rm tr}(F\wedge F) + (b){\rm tr}(R \wedge R) \sim {\bar \partial} H
\eeq
for some constants $a, b$ and closed three-form $H \in H^1(TX^{\vee} \otimes TX^{\vee})$. Thus, it is clear  that in heterotic theories with a bundle ${\cal P}$ satisfying the anomaly cancellation condition \eref{bi1} it is always possible to define a transitive Courant algebroid of the form ${\cal U}$ in \eref{the_goal} satisfying $p_1({\cal U})=\langle F_{total}, F_{total} \rangle$ for $F_{total}$ the field strength associated to \eref{vtot} and some choice of the pairing $\langle , \rangle$ \cite{Baraglia:2013wua}. It is important to recall here that the Pontryagin class in this context is defined {\it for each} choice of pairing $\langle , \rangle$ \footnote{In the case that $G=GL_n(\mathbb{C})$ and the pairing is given by the trace of product matrices, the Pontryagin class is $2 \textnormal{ch}_2$.}. See \cite{Baraglia:2013wua}, Proposition 3.2 for the details on the definition of the bracket, and so forth in the case of a heterotic transitive Courant algebroid.

In summary then, we see that the very short exact sequences defined in Section \ref{Sec:seq}:
\begin{align}
& 0 \to TX^{\vee} \to {\cal H} \to {\cal Q} \to 0 \;,\label{courant_final1} \\
& 0 \to \textnormal{End}_0(V)\oplus \textnormal{End}_0(TX) \to {\cal Q} \to TX \to 0 \;,\label{courant_final2}
\end{align}
are actually the dual of a heterotic Courant algebroid -- a transitive Courant algebroid built out of the principal bundle ${\cal V}_{total}={\cal V}_{gauge} \oplus {\cal V}_{frame}$, defined if and only if the anomaly cancelation condition is satisfied. It is a remarkable correspondence  that the short exact sequences which arose purely from the structure of the infinitesimal fluctuation of the Strominger system also arise naturally in the rich mathematical subject of Courant algebroids. As we will see in the following sections, it may be that this correspondence hints at deeper links between non-K\"ahler heterotic geometries and algebroids arising in generalized geometry. To begin, we next consider an even simpler origin for the transitive Courant algebroids arising in heterotic theories.

\subsection{Algebroids by reduction}

\subsubsection{Atiyah algebroids by reduction}\label{reduction_atiyah}
We begin by recalling a familiar and elegant story -- the derivation of the Atiyah algebroid from that of the familiar deformation space of a compact manifold. Recall that in the case of the Atiyah algebroid, the origin of the short exact sequence \eref{atiyah} could be straightforwardly understood in terms of infinitesimal complex deformations of the total space of a vector bundle. Following \cite{donaldson} recall that the simultaneous deformation space, $Def(X,V)$, of a vector bundle $V \to X$ and its base manifold can be reduced to the familiar case of  complex deformations of a compact manifold by treating the well-understood deformations of the line bundle $\wedge^{{\rm max}}(V)$ and the projective bundle $\mathbb{P}_{V}=\mathbb{P}(V) \stackrel{r}{\longrightarrow} X$ separately. In this case there is an exact sequence of sheaves
\beq
0 \to \Theta_{\mathbb{P}_{V}|X} \to T{\mathbb{P}_{V}} \to r^{*}(TX) \to 0
\eeq
on the projectivized total space $\mathbb{P}_{V}$, where $\Theta_{\mathbb{P}_{V}|X}$ denotes the vertical vector fields. The reduction of this sequence to $X$ leads to the familiar Atiyah sequence \eref{atiyah} and moreover, the Leray spectral sequence for $r$ gives the relationship
\beq\label{reduc_atiyah_def}
\cdots
\to H^1(X, \textnormal{End}_0(V)) \to H^1({\mathbb{P}_{V}},T{\mathbb{P}_{V}}) \to H^1(X, TX) \to H^2(X, \textnormal{End}_0(V)) \to
\cdots \;.
\eeq
This is the statement that the usual deformations of ${\mathbb{P}_{V}}$ as a compact, complex manifold reduce to the description of the Atiyah deformations $H^1(X, {\cal Q})$ in \eref{thirdone} of the pair $(X,V$) built from the deformations of $V$ (with $X$ fixed) and the deformations of $X$ (regardless of $V$).

As we will see below, this ``reduction" structure is remarkably similar to that which occurs for transitive Courant algebroids in heterotic theories. As shown in \cite{Baraglia:2013wua, garcia_ferndandez}, heterotic Courant algebroids can be obtained by reduction of {\it exact Courant algebroids}.

\subsubsection{Exact Courant algebroids on the total space of a bundle}\label{exact_algebroid}
A Courant algebroid, $E$ is called {\it exact} if $E$ is transitive and the kernel of the anchor map coincides with the image of the map $\rho^{*}: TX^{\vee} \to E$. Since $E$ is transitive the map $\rho^{*}$ is injective and hence $TX^{\vee}$ is a sub-bundle of $E$. This leads to the short exact sequence
\beq\label{exact_total}
0 \to TX^{\vee} \stackrel{\rho^{*}}{\longrightarrow} E \stackrel{\rho}{\longrightarrow} TX \to 0 \;.
\eeq
In the case at hand, it has been shown that heterotic Courant algebroids arise via reduction of an exact Courant algebroid on the total space of the principal bundle. In general, for any $G$-principal bundle $\sigma: {\cal P} \to X$ an exact Courant algebroid can be constructed that is a non-trivial extension of $T{\cal P}^{\vee} \oplus T{\cal P}$ characterized by a $G$-invariant $3$-form $H \in \Omega^3(P)$ \cite{bressler1}.

According to \cite{Baraglia:2013wua, garcia_ferndandez} (see Proposition 3.5 in \cite{Baraglia:2013wua} and Section 2 of \cite{garcia_ferndandez}) every heterotic Courant algebroid of the form \eref{courant_alg1}-\eref{courant_alg2} on $X$ is obtained by reduction of an exact Courant algebroid, $E$ of the form \eref{exact_total} on the principal bundle ${\cal P}_{total}={\cal P}_{gauge} \oplus {\cal P}_{frame}$ with $G$-invariant three-form $H \in \Omega^{3}({\cal P}_{total})$. Given a class $h=[H] \in H^3({\cal P}_{total})$ it is possible to reduce the Courant algebroid and reproduce the familiar geometric ingredients (gauge connection, three-form flux, and so forth) on $X$. The reduction follows the procedure of \cite {bcg} is achieved by an ``extended action" $\xi: {\mathfrak g} \to \Gamma(T{{\cal P}_{total}}^{\vee})$ of the form. 
\beq
d_{G}(H+ \xi)= \langle, \rangle \;,
\eeq
where $d_G$ is the differential of the Cartan complex (see \cite{Baraglia:2013wua} for details). In  \cite{Baraglia:2013wua} it is shown that this reduction can uniquely be achieved by
\beq
\xi=-\langle , \rangle A_{tot}~,
\eeq
where $A_{tot}$ is a connection on ${\cal P}_{total}$ such that the three form $H$ on ${\cal P}_{total}$ is defined by
\begin{align}\label{fiber1}
&H=\sigma^{*}(H_0)-CS_3(A_{tot})\;, \\
&CS_{3}(A_{tot})=\langle A_{tot}, F_{tot} \rangle -\frac{1}{3!} \langle A _{tot}, [A_{tot},A_{tot}] \rangle \;, \label{fiber2}
\end{align}
with $H_0$ a 3-form on $X$ satisfying $dH_0 = \langle F_{tot}, F_{tot} \rangle  \sim \textnormal{tr} (R \wedge R) - \textnormal{tr}(F \wedge F)$, i.e. the anomaly cancellation condition \eref{bi1}. Of course to fully define the Courant algebroids we must also define the relevant brackets. These can be found in Appendix \ref{Courant_app}. The form of \eref{fiber1}-\eref{fiber2} demonstrates that extended actions reducing exact Courant algebroids to heterotic transitive Courant algebroids exist if and only if the class $h \in H^3({\cal P}_{total})$ is such that its restriction to the fibers of ${\cal P}_{total}$ coincides with the relevant Cartan $3$-forms $\omega_3 \in H^3(G_{total}, \mathbb{R})$ determined by the pairing $\langle , \rangle$. 

Furthermore, upon restricting the threeform in \eref{fiber1} to the base manifold, it is clear that it plays the role of the defining extension class (and associated co-boundary map in cohomology) of \eref{courant_alg1}, \eref{courant_alg2}. That is, it is of the form $H_0 - \omega_3^{total}$ satisfying the anomaly cancellation condition. Moreover, by inspection we see that this is exactly of the form of the map $M$ in \eref{mdef}:
\beq
M = \partial J + i \frac{1}{30} \alpha' \omega_3^{\textnormal{YM}} - i \alpha' \omega_3^{\textnormal{L}} + M^0 \;,
\eeq
derived from the fluctuation analysis of Sections \ref{Sec:perturb} and \ref{Sec:seq}. As hoped, the short exact sequences \eref{atiyah}, \eref{groupoid} defining a transitive Courant algebroid derived via the fluctuation analysis and by the study of generalized geometry agree!

With the important observation in hand that the transitive Courant algebroids arising in heterotic theories naturally descend from simpler exact Courant algebroids on ${\cal P}_{total}$ we can now compare this structure to the case of Atiyah algebroids in Section \ref{reduction_atiyah}. In the Atiyah algebroid case we were able to relate a complicated simultaneous deformation problem (that of $(X,V)$) to the simpler problem of complex deformations of a compact manifold: ${\mathbb P}_{total}$. In the present case, it is possible to ask the same question, namely do the infinitesimal holomorphic deformations measured by $H^1({\cal H})$ in \eref{cohfluct} described in Section \ref{Sec:seq} descend from some simpler deformation problem on ${\cal P}_{total}$? In the next Section we will see that the answer to this question leads us away from ordinary complex deformation theory into the realm of Hitchin's generalized geometry \cite{first_hitchin,hitchin05}. While Hitchin's generalized geometry has been successfully utilized in compactifications of Type II string theories and M-theory (see \cite{Grana:2004bg,Grana:2005sn} for example), its applicability to heterotic string theory has remained an open question. We shall see that the structures/deformations explored in this work hint at exactly such a connection.

\subsection{Links to Hitchin's generalized geometry}\label{gen_geom_sec}
Courant algebroids play a fundamental role in the subject of Hitchin's generalized complex structures \cite{first_hitchin,gualt_thesis}. In one of several equivalent definitions, we can define a generalized complex structure on a manifold $X$ as an almost complex structure ${\cal J}$ on the exact (split) Courant algebroid $E=TX \oplus TX^{\vee}$ which is orthogonal with respect to the pairing $\langle , \rangle$ (this is a reduction of the structure of the $O(2n,2n)$-bundle $TX \oplus TX^{\vee}$ to the group $U(n,n)$).

Much like in the case of ordinary complex structures on a space $X$ -- with infinitesimal deformation space defined as the kernel of a map $\Phi: H^1(X, TX) \to H^2(X, TX)$ -- it is possible to define the infinitesimal deformations of a generalized complex structure. From Gualtieri's thesis (\cite{gualt_thesis}, Theorem $5.4$) it is known that the infinitesimal deformation space of a generalized complex structure is contained in an open neighborhood\footnote{Here $L$ is the $+i$-eigenbundle of the generalized complex structure ${\cal J} \in O(TX \oplus TX^{\vee})$ and there is a differential graded algebra $(\wedge^{\bullet}L^{\vee}, d_{L})$. See \cite{gualt_hodge} for details.} in $H^{2}_{L}(X)$ with obstructions in $H^{3}_{L}(X)$.

It is beyond the scope of the present paper to consider the reduction of $H^{2}_{L}(X)$ under a Leray-type spectral sequence, but it is tempting to speculate that (as in the Atiyah case, Section \ref{reduction_atiyah}, \eref{reduc_atiyah_def}) this infinitesimal deformation space is related to $H^2(X, {\cal U})$ and we intend to explore this in future work. Moreover, it should be recalled that under Serre duality $H^2(X, {\cal U}) \simeq H^1(X, {\cal H})$. That is, a fluctuation of the generalized complex structure on the total space of ${\cal P}_{total}$ could lead to the first order deformation space, $H^1(X, {\cal H})$, given in \eref{cohfluct} and Section \ref{Sec:seq}, describing the infinitesimal deformations of the Strominger system as in Section \ref{Sec:perturb}.

Carrying this speculation a step further, it may be possible to link the full moduli of the non-K\"ahler Strominger system to generalized geometry. One suggestive hint in this direction was provided in recent work \cite{garcia_ferndandez}. Here it was shown in the context of $10$-dimensional, flat-space heterotic supergravity, that the exact Courant algebroid \eref{exact_total} defined on ${\cal P}_{total}={\cal P}_{gauge} \oplus {\cal P}_{frame}$ can also be endowed with generalized K\"ahler structures and a generalized metric. According to \cite{garcia_ferndandez} an admissible generalized metric on $E$ satisfies
\beq
GRic=0
\eeq
(the vanishing of the generalized Ricci-tensor) if the underlying heterotic fields on the base manifold $\mathbb{R}^{1,9}$ satisfy the equations of motion of heterotic supergravity in $10$-dimensions. With better understanding of spinors in transitive Courant algebroids it might be possible to combine these $10$-dimensional results with those of \cite{Baraglia:2013wua} for compactifications in order to fully explore Hitchin's generalized geometry  and its deformations in the context of the Strominger system. We hope this intriguing topic will be developed in the future.

For now, having reviewed the subject of Courant algebroids as they arise in the context of this work and the Strominger system, we move away from the supergravity limit to discuss these same structures as they appear in heterotic sigma models.

\section{Relationship to $(0,2)$ NLSMs and the $\alpha'=0$ limit}\label{Sec:sigma}
The problem of understanding moduli in perturbative
heterotic compactifications on
possibly non-K\"ahler manifolds was studied in
\cite{ms} by enumerating BRST-closed operators in heterotic
nonlinear sigma models.

For Calabi-Yau (0,2) compactifications, since one can go to a weak-coupling limit of the nonlinear sigma model, this method is on solid grounds, and reproduces existing results on the role of Atiyah classes \cite{Anderson:2010mh,Anderson:2011ty,Anderson:2013qca}.  For non-K\"ahler compactifications, this method is necessarily more formal, as one cannot smoothly deform to a weak coupling large-radius limit.  Nevertheless, the analysis applies to a reasonable approximation in cases where the compactification curvature is large compared to the string scale. As we shall outline here, the sigma model computation dovetails with the structure we have seen by perturbing supergravity, in the limit where one take $\alpha'=0$.

Let us quickly review the worldsheet results.
Briefly, the paper \cite{ms} wrote down the most general possible
supersymmetric marginal operator deforming the classical action.
In (0,2) superspace following \cite{ms}, this had the form $D {\cal O}$ for
${\cal O}$ a superfield annihilated by $\overline{D}$, with classical dimension
1 and $U(1)_R$ charge $+1$.  The most general operator satisfying the
second two conditions
is of the form
\begin{equation}
{\cal O} \: = \: 
\left[ \overline{\Gamma}_{\alpha} \Gamma^{\beta} \Lambda^{\alpha}_{\beta 
\overline{a}}(\Phi, \overline{\Phi}) \: + \:
\partial \Phi^a Y_{a \overline{a}}(\Phi, \overline{\Phi}) \: + \:
\partial \overline{\Phi}^{\overline{b}} g_{a \overline{b}}
Z^a_{\overline{a}}(\Phi, \overline{\Phi}) \right]
\overline{D} \overline{\Phi}^{\overline{a}},
\end{equation}
where $\Gamma$'s are Fermi superfields coupling to the gauge bundle,
$\Phi$'s are chiral superfields describing the right-moving degrees of
freedom, and $Z^a_{\overline{a}}$, $Y_{a \overline{a}}$, and
$\Lambda^{\alpha}_{\beta 
\overline{a}}$ are bundle-valued differential forms defining the deformations.  
Demanding that $\overline{D} {\cal O} = 0$ gives cocycle conditions 
\begin{align}
Z^a_{\overline{b}, \overline{c}} \: - \: 
Z^a_{\overline{c}, \overline{b}} & ~=~ 
0,\label{eq:cocycle1} \\
Y_{a \overline{b}, \overline{c}} \: - \: Y_{a \overline{c}, \overline{b}}
& ~=~  
Z^b_{\overline{c}} H_{b a \overline{b}} \: - \:
Z^b_{\overline{b}} H_{b a \overline{c}}, \label{eq:cocycle2} \\
\Lambda^{\alpha}_{\beta \overline{a}, \overline{b}} \: - \:
\Lambda^{\alpha}_{\beta \overline{b}, \overline{a}} &~ =~ 
F^{\alpha}_{\beta \overline{b} a} Z^a_{\overline{a}} \: - \:
F^{\alpha}_{\beta \overline{a} a} Z^a_{\overline{b}}.
\label{eq:cocycle3}
\end{align}

In passing, we can identify the cocycle conditions above with the
$\alpha' \rightarrow 0$ limit of the
moduli conditions given earlier in equations (\ref{one}) and (\ref{two}).
Specifically, in this limit, the cocycle condition~(\ref{eq:cocycle2})
corresponds precisely to (\ref{one}), if we identify 
\begin{equation}
Z^a_{\overline{b}} \: = \: \delta J^a_{\overline{b}}, \: \: \:
H \: = \: \partial J, \: \: \:
Y_{a \overline{b}} \: = \:
i \delta J_{\overline{b} a} \: - \: 2 \delta B_{\overline{b} a} \: - \:
\Lambda^{\alpha'}_{\overline{b} a}.
\end{equation}
Holomorphicity of $\delta J^a_{\overline{b}}$ corresponds to the
cocycle condition~(\ref{eq:cocycle1}), and finally the last cocycle
condition~(\ref{eq:cocycle3}) above corresponds to (\ref{two}), and
gives a local description of the Atiyah sequence \cite{atiyah}.

Unlike the computations in this paper,
the worldsheet analysis of \cite{ms} does not overcount moduli, as it does
not distinguish spin connection deformations from metric deformations, 
but in the limit $\alpha' 
\rightarrow 0$ where such decouple, we see that the results of this paper
effectively match those of \cite{ms}.

As a more technical aside,
the reader should also note that worldsheet analyses such as the
above identify BRST-closed states with massless low-energy states,
whereas the supergravity analysis described earlier in this paper will,
in general, describe additional states albeit with couplings that will
generate masses.  In effect, the results of a worldsheet analysis
will only match the enumeration of massless states in a supergravity
analysis after integrating out massive degrees of freedom.

Working on the worldsheet, it is also possible to derive coboundaries. If two marginal operators differ by a superspace derivative, they define
the same deformation.  Similarly, contributions such that $D{\cal O}$ is a
total derivative leave the theory unchanged.
Such identifications result in coboundaries of the
form
\begin{align}
Z^a_{\overline{b}} & ~\cong ~ Z^a_{\overline{b}} \: + \: 
\left(\zeta^a + g^{a \overline{c}} \xi_{\overline{c}} \right)_{, \overline{b}}
\: + \:
g^{a \overline{c}} \left( \xi_{\overline{b}, \overline{c}} - 
\xi_{\overline{c},\overline{b}} \right), \\
Y_{a \overline{b}} & ~\cong ~ Y_{a \overline{b}} \: + \:
\mu_{a, \overline{b}} \: + \: \xi_{\overline{b}, a} \: + \:
H_{a \overline{b} c} \left( \zeta^c + g^{c \overline{d}} \xi_{\overline{d}}
\right), \\
\Lambda^{\alpha}_{\beta \overline{a}} &~ \cong ~
\Lambda^{\alpha}_{\beta \overline{a}} \: + \:
\lambda^{\alpha}_{\beta, \overline{a}} \: - \:
F^{\alpha}_{\beta \overline{a} b} \left( \zeta^b + g^{b \overline{c}} 
\xi_{\overline{c}} \right).
\end{align}

On the (2,2) locus, it can be shown \cite{ms} that the cocycle conditions
can be simplified, and then $Z$, $Y$, and $\Lambda$ can be interpreted
as complex, K\"ahler, and bundle moduli, respectively.

For Calabi-Yau (0,2) theories, where at leading order $H=0$,
the data above defines complex and K\"ahler moduli, plus bundle moduli
that are intertwined with K\"ahler moduli as described by the
Atiyah sequence \cite{Anderson:2010mh,Anderson:2011ty,Anderson:2013qca}.

The results above also have something to say about non-K\"ahler (0,2)
theories, with the important caveat that as this is a worldsheet NLSM
computation, it is implicitly only reliable near large-radius (small
$\alpha'$) limits.  Given that limitation, the cocycle 
condition~(\ref{eq:cocycle2}) indicates
an Atiyah-like structure mixing the complex and K\"ahler moduli.
If we, formally, take $\alpha'=0$, so that $H$ is a closed form,
then the (2,1) part of $H$ defines an element of
\begin{equation}
H^1(\wedge^2 TX^{\vee}) \: \subseteq \: H^1(TX^{\vee} \otimes TX^{\vee})
\end{equation}
and hence an extension
\begin{equation}   \label{eq:C-exact}
0 \: \longrightarrow \: TX^{\vee} \: \longrightarrow \: {\cal E} \:
\longrightarrow \: TX \: \longrightarrow \: 0~.
\end{equation}
where the complex and K\"ahler moduli are replaced by $H^1({\cal E})$. Thus, in the context of NLSMs and the $\alpha'=0$ limit, we are once again naturally lead to Courant algebroids. At this leading order, we find an exact Courant algebroid on $X$ itself (rather than ${\cal P}_{total}$) as defined in Section \ref{exact_algebroid}.

Given such an exact Courant algebroid in the smooth category we can look at the 
a curvature 3-form $H \in \Omega^3_{cl}(X)$ in more detail, following
\cite{bcg}.  In the smooth category, the exact sequence above splits, so let $\nabla: TX \rightarrow
{\cal E}$ be a splitting whose image in ${\cal E}$ is isotropic with respect to $\langle , \rangle$.  Then, for $v, w$ tangent vectors to
$X$, $H$ is defined by
\begin{displaymath}
i_w i_v H \: = \: 2 s [ \nabla(v), \nabla(w) ],
\end{displaymath}
where $s: {\cal E} \rightarrow TX^{\vee}$ is the induced left splitting.
It can be shown that different choices of splittings change $H$ by
$dB$ for some 2-form $B$.  This important observation highlights the type of deformation problem that this exact Courant algebroid is describing: when $H$ represents an element of integral
cohomology, the corresponding exact Courant algebroid can be viewed as an
analogue of an Atiyah sequence, but for connections on $U(1)$ gerbes. 

Furthermore, given an exact Courant algebroid as above,
we can put a Courant algebroid
structure on $TX \oplus TX^{\vee}$.  Given $v + \xi, w + \eta \in \Gamma( TX
\oplus TX^{\vee})$, we define
\begin{eqnarray*}
\langle v + \xi, w + \eta \rangle & = & \frac{1}{2} \left( \eta(v) + \xi(w) 
\right) , \\
\left[ v + \xi, w + \eta \right]_H & = & [v,w] \: + \: {\cal L}_v \eta \: - \:
i_w d \xi \: + \: i_w i_v H ,
\end{eqnarray*}
where the $[,]_H$ above is the $H$-twisted Courant bracket on $TX \oplus
TX^{\vee}$. So far we have discussed Courant algebroids in the smooth category.
Holomorphic exact Courant algebroids have been discussed in \cite{gualt10}.
These have a characteristic class in $H^1(\Omega^{2,cl}(X))$,
which can be understood as classifying extensions of $TX$ by $TX^{\vee}$,
see for example
\cite{gualt10} ({\it e.g.} examples 1.1, 1.4),
or \cite{hitchin05} (Section 2.3) for a construction of $Q$ from a $B$ field associated to the pertinent gerbe.

With these observations in hand, we can return to the definition of \eref{eq:C-exact} and understand the moduli of \eref{eq:cocycle1}- \eref{eq:cocycle3} in terms of this exact Courant algebroid. In the special case that $H=0$, ({\it e.g.} $X$ is a Calabi-Yau manifold to leading order)
then ${\cal E} = TX^{\vee} \oplus TX$, and 
\begin{displaymath}
H^1({\cal E}) \: = \: H^1(TX^{\vee}) \oplus H^1(TX),
\end{displaymath}
so that we recover the usual complex and K\"ahler moduli. Returning to equation~(\ref{eq:C-exact}), it is natural to consider $H^1({\cal E})$ (albeit this is only meaningful in a formal $\alpha' \rightarrow 0$ limit).  To do this, we can use the
associated long exact sequence,
\begin{displaymath}
H^1(TX^{\vee}) \: \longrightarrow \: H^1({\cal E}) \: \longrightarrow \:
H^1(TX) \: \longrightarrow \: H^2(TX^{\vee}),
\end{displaymath}
where the coboundary map will be given by contraction with
$H$. Taking the $H^0(TX)=0$ case for simplicity it is clear that 
\beq
H^1({\cal E}) = H^1(TX^{\vee}) \oplus \textnormal{ker}(H^1(TX) \: \longrightarrow \: H^2(TX^{\vee})).
\eeq
This is exactly the cocycle condition, \eref{eq:cocycle1}, derived in \cite{ms}.

It is clear that this will lead to only a subset of the K\"ahler/complex structure. As a trivial consistency check of this result in the $H \neq 0$ case, it can be noted that
one of the few things generally acknowledged about moduli of non-K\"ahler
heterotic compactifications is that the overall K\"ahler `breathing' mode,
rescaling the entire metric by a factor, is obstructed.
To that end, note that in cocycle condition~(\ref{eq:cocycle2}),
if we take $Z=0$ and $Y_{a \overline{b}} \propto g_{a \overline{b}}$
(so as to describe the breathing mode), then since the space is
non-K\"ahler, $\overline{\partial} Y \neq 0$, so the cocycle condition is
not obeyed, and the breathing mode is obstructed.

Putting together the results of the NLSM analysis and the sequences and cohomology analyzed thus far, the short exact sequence defining an extension ${\cal E}$
whose degree one cohomology describes the pertinent subset of complex and
K\"ahler moduli, is seen to precisely coincide with a holomorphic exact Courant algebroid. Recall once again that Courant algebroids describe deformations of coupled structures. Here the exact Courant algebroids above encode infinitesimal symmetries of the ${\mathbb C}^{\times}$ gerbe characterized by the characteristic class in $H^1(\wedge^2 TX^{\vee})$ 
and `compatible' complex structure deformations of $X$. See \cite{collier11} for related information on symmetries of exact Courant algebroids.

To discuss how one would actually compute these deformation spaces, even in this $\alpha'=0$ limit, we must consider more explicitly how to define the co-boundary map $H$,
\beq
H^1(TX) \stackrel{H}{\longrightarrow} H^2(TX^{\vee}).
\eeq
There is a close analogue to the structure above which arises in Noether-Lefschetz theory \cite{katz-priv}.
Let $S$ be a K3, and $C \subset S$ a curve.
$[C] \in H^1(TS^{\vee})$, and $[Z] \in H^1(TS)$.
The pairing
\begin{displaymath}
\varphi: \: H^1(TS) \otimes H^1(TS^{\vee}) \: \longrightarrow \: H^2({\cal O}_S)
\end{displaymath}
determines whether $C$ deforms holomorphically under the complex structure
modulus $Z$ -- it does, if and only if $\varphi([Z]\otimes[C]) = 0$.
Since $H^2({\cal O}_S)$ is one-dimensional, and since this pairing is
nondegenerate, this imposes one constraint equation, eliminates one
degree of freedom. For example, the space of (generically nonalgebraic) K3's is 20-dimensional,
but if we demand that a curve be holomorphic,
then we get a 19-dimensional moduli space, and $20-1=19$.

In the present circumstances, we have a higher-form analogue of
Noether-Lefschetz theory.  $[H] \in H^1(\wedge^2 TX^{\vee} )$ and
$[Z] \in H^1(TX)$, so the pairing $H \cdot Z$ defines a map
\begin{displaymath}
H^1(TX) \otimes H^1(\wedge^2 TX^{\vee}) \: \longrightarrow \: 
H^2(TX^{\vee}).
\end{displaymath}
From linear algebra, this can impose up to $h^2(TX^{\vee})$ constraints,
depending upon the degeneracy of the pairing.
On a threefold with $K_X$ trivial, by Serre duality\footnote{
See {\it e.g.} \cite{btt,bs1} for a discussion of Serre duality and
Riemann-Roch on non-K\"ahler manifolds.
},
$h^2(TX^{\vee}) = h^1(TX)$, hence there are potentially as many constraints as elements of $H^1(TX)$.

As a final note we remind the reader that at this order in $\alpha'$, 
the final two cocycle conditions \eref{eq:cocycle2} and \eref{eq:cocycle3} are 
simply the de-coupled Atiyah sequences describing the holomorphic deformations 
of $V$ and $TX$. Thus, in complete agreement with the $\alpha'=0$ limit of the 
results of Section \ref{well_defined}, 
we have seen that the leading order moduli correspond to those arising from a Courant algebroid.

\section{Conclusions and future work} \label{discussion}

In this paper we have studied metric, spin connection, and bundle
moduli of K\"ahler and non-K\"ahler heterotic
string compactifications through first order in $\alpha'$
via low-energy supergravity deformations.
We have recovered the heterotic non-K\"ahler moduli obtained in
\cite{ms} at $\alpha'=0$ as a special limit.  

For $\alpha'\neq0$
our methods produce a potentially redundant description of the physical moduli, in which the D-flatness conditions (\ref{d1}) have not yet been imposed. In addition, the metric and spin
connection deformations are distinguished, leading to a potential overcounting in these degrees of freedom. The result
has a tantalizingly simple understanding as the cohomology group
$H^1({\cal H})$, where ${\cal H}$ is a bundle extension obtained by
{\it e.g.} \cite{Baraglia:2013wua} as part of an otherwise-unrelated
realization of the heterotic anomaly cancellation condition in the
language of Courant algebroids.

It is important to note that the results of this paper hold only for heterotic compactifications on non-K\"ahler manifolds satisfying the $\partial \bar{\partial}$-lemma. However, there are many known examples of Strominger system  compactifications on such spaces, including
the well-known ``non-standard embeddings" (deformations away from Calabi-Yau threefolds), as well as ``fully non-K\"ahler" possibilities (for example \cite{Fu:2008zh} and some of the geometries in \cite{Goldstein:2002pg,Fu:2006vj}). In a future publication, we hope to apply the formalism we have developed here to such examples. 

Our results lead to a number of natural and intriguing questions that it would be illuminating to explore in the future. These include the following questions and future directions:

\begin{itemize}
\item Apply the formalism developed here to explicitly compute $H^1({\cal H})$ on examples of non-K\"ahler compactifications satisfying the $\partial\bar{\partial}$-lemma.
\item Extend the analysis of this work to include constraints from the ``D-term" conditions (\ref{d1}) and explicitly determine the redundancy in the parameterization of the ``F-flat" deformation space described by $H^1({\cal H})$. With the results of the current work and these next steps in hand, it would be possible to explicitly determine the full infinitesimal moduli space of the heterotic Strominger system. 
\item Determine the relationship between $H^1({\cal H})$ and the deformations of a generalized complex structure on $\mathbb{P}({\cal V}_{total})$ as conjectured in Section \ref{gen_geom_sec}.
\item It would naturally be of great interest to be able to generalize these results to non-K\"ahler compactifications which do not satisfy the $\partial \bar{\partial}$-lemma. However, as pointed out in Section \ref{intro}, there are a number of manifest difficulties which arise immediately, including the fact that relevant operators are no longer elliptic and infinitesimal deformations of the conformally balanced metric need no longer be balanced. Despite this, some progress has been made in determining the moduli of such non-K\"ahler compactifications in the context of Type II theories \cite{Tseng:2011gv} and we hope that in future such results may be extended to the heterotic context. 
\end{itemize}

While the primary motivation of this work was to develop new tools and the formalism to understand heterotic non-K\"ahler compactifications, the significance of these results for more familiar compactifications should not be overlooked. We conclude by briefly putting our results in context for non-K\"ahler deformations of smooth heterotic Calabi-Yau compactifications and considering the implications for heterotic non-standard embeddings and string phenomenology. Such deformations of Calabi-Yau backgrounds are an important and simple class of non-K\"ahler compactifications which satisfy the $\partial \bar{\partial}$ lemma.

The first compactifications of the heterotic string were the so-called ``standard embeddings"  \cite{candelas-horowitz} in which the gauge bundle $V$ is taken to be the holomorphic tangent bundle to a Calabi-Yau threefold. Despite the simplicity of such Calabi-Yau geometries, the search for heterotic compactifications that could be relevant for string phenomenology -- {\it i.e.} produce Standard Model type gauge theories and particle spectra -- naturally led to the consideration of other, non-standard embeddings \cite{Witten:1985bz}. In these, the vector bundle $V$ was chosen to have a higher rank structure group ($SU(4)$ or $SU(5)$ for example), leading to more physically relevant $4$-dimensional $SO(10)$ or $SU(5)$ gauge theories that could be broken to the Standard Model. However, this phenomenological progress comes with a well-known increase in mathematical complexity. A non-standard embedding deforms the background geometry away from Ricci-flat K\"ahler to higher order in $\alpha'$ \cite{Witten:1985bz, Witten:1986kg}. Such solutions to the heterotic equations of motion were shown explicitly to exist in \cite{Li:2004hx}.

By working to first order in $\alpha'$, the non-K\"ahler nature of the background geometry can in many ways be effectively ignored. However, to fully address the problem of moduli stabilization, it is important to understand the coupled fluctuation problem described in this work.

As first explored in \cite{Anderson:2010mh,Anderson:2011ty,Anderson:2013qca}, the understanding of the actual deformation moduli of a general Calabi-Yau compactification with $V \neq TX$ is an important tool in the problem of moduli stabilization. Indeed, it was demonstrated that by considering the simultaneous Atiyah deformation space $Def(X,V)$, that the number of physical moduli of the effective theory could be far fewer than the naive count $h^1(TX) + h^1(TX^{\vee}) + h^1(End_0(V))$. In certain regions of moduli space this reduction of the naive moduli fields was shown to be describable as F-term lifting through a Gukov-Vafa-Witten super potential
\beq\label{gvw}
W \sim \int_{X} H \wedge \Omega^{3,0}~.
\eeq
Furthermore, it was shown \cite{Anderson:2011cza} that choosing vector bundles which were only holomorphic for higher co-dimensional loci of their base manifold, $X$, and slope-stable only for sub-cones of K\"ahler moduli space, could in principle fix all but one of the {\it geometric} moduli of a heterotic Calabi-Yau compactification. However, it was also clear that such perturbative moduli stabilization scenarios were still incomplete, since for example, the structure of the Atiyah deformation $Def(X,V)$ space did not constrain the vector bundle moduli $H^1(X,\rm{End}_0(V))$.

In this work, we have extended the analysis of the coupled holomorphic deformation problem $Def(X,V)$ to include the heterotic three-form. That is, we are considering an analogous holomorphic deformation of the triple $Def(X,V,H)$. From the results of Section \ref{Sec:seq} it is clear that this simultaneous deformation problem can in principle remove even more moduli from a heterotic compactification at higher orders in $\alpha'$. For example, by comparing the dimension sources/targets in \eref{order1}, it is clear that in principle more of the naive deformations lying in $H^1(TX)$ and $H^1(\textnormal{End}_0(V))$ could be obstructed. 

Finally, in the context of F-term conditions in Calabi-Yau compactifications, it would be good to understand the relationship of these effects to known higher order (Kuranishi) obstructions arising in the deformation theory and their appearance in the super potential \eref{gvw} (see \cite{Anderson:2011ty} for a discussion). We hope to explore the physical consequences of this deformation theory and the role of $H^1({\cal H})$ in future work.

\section*{Acknowledgements}

We would like to thank A.~Caldararu, M.~Gualtieri, S.~Katz, I.~Melnikov and S.-T. Yau for useful discussions.
E.S. was partially supported by NSF grant PHY-1068725.

\appendix
\section{Some details on Courant algebroids}\label{Courant_app}
In this Appendix we include a few standard definitions for completeness. The definitions below are taken from the nice review \cite{shlomo} and we follow the conventions/notation laid out there.

\subsection{Groupoids, algebroids, and so forth}
\begin{definition}
A {\it Groupoid}, $\mathfrak{G}$, is a (small) category in which every arrow is invertible. A groupoid has a base manifold, $X$ and $\mathfrak{G}$ is said to be a ``groupoid over $X$." 
\end{definition}
An elementary but illustrative example of a groupoid is given by the set of all linear isomorphisms from one fiber to another of a vector bundle $V \to X$. In addition, any principal $G$-bundle ${\cal P} \to X$ has a so-called ``gauge groupoid", whose objects are points of $X$, and whose morphisms are elements of the quotient of ${\cal P} \times {\cal P}$ by the diagonal action of $G$, with source and target morphisms given by the two projections of $X$. An infinitesimal version of a smooth groupoid is a Lie algebroid:

\begin{definition}
Let $X$ be a smooth manifold. A {\it Lie algebroid} is a vector bundle $V$ over $X$ where $q: V \to X$, together with a bundle map, $\rho: X \to TX$ called the {\it anchor} and a bracket
\beq
[ , ]: \Gamma(V) \times \Gamma(V) \to \Gamma(V)
\eeq
which is skew-symmetric, bilinear and satisfies the Jacobi identity (and so makes $\Gamma(V)$ into a Lie algebra) subject to the axioms
\begin{align}
& [U, fW]=f[U,W] + (\rho(U)f)W, \\
&\rho([U,W])=[\rho(U), \rho(W)]~,
\end{align}
where $U,W \in \Gamma(V)$ and $f \in C^{\infty}(X)$
\end{definition}
There are other equivalent definitions of Lie algebroids including a differential operator on sections of $\wedge V^*$ and in terms of Poisson structures (see \cite{shlomo,chen}). The Atiyah algebroid, defined in Section \ref{Sec:algebroids} associated to a principal $G$-bundle ${\cal P}$ (where $G$ is a Lie group) is the Lie algebroid of the gauge groupoid of ${\cal P}$.

Finally, as defined in Section \ref{Sec:algebroids}, a Courant algebroid is a Lie algebroid with the additional
structure of a fiber-wise inner product:

\begin{definition}
A Courant algebroid consists of the following: a vector bundle $V \to X$ over a smooth manifold, $X$, a bilinear operator $[,]: \Gamma(V) \otimes \Gamma(V) \to \Gamma(V)$ on the space of sections of $V$, a non-degenerate bilinear form, $\langle , \rangle$ on $V$ and an anchor map $\rho: V \to TX$. The data $(V, \rho,[,], \langle , \rangle )$  is called a ``Courant algebroid" if the conditions below hold for all $a,b,c \in \Gamma(V)$. A Courant algebroid is called ``regular" if the anchor map is of constant rank. The bracket $[,]$ can be either symmetric or skew-symmetric.\footnote{Referred to in the literature as the ``Dorfman" or ``Courant" bracket respectively, though both can arise as Courant brackets above.}
\begin{itemize} 
\item $[a, [b,c]]=[[a,b],c]+[b,[a,c]]$,
\item $[a,b] + [b,a]=2d \langle a, b \rangle$,
\item $\rho(a) \langle b, c \rangle = \langle [a,b],c \rangle + \langle b, [a,c] \rangle$.
\end{itemize}
\end{definition}

As a consequence of the definitions above, Courant algebroids satisfy not only Jacobi-type identities and but also Liebniz rules:
\begin{align}
& [a, fb]=f[a, b] + \rho(a)(f)b, \\
&\rho[a, b] = [\rho(a), \rho(b)].
\end{align}

\begin{definition}
$V$ is called ``transitive" if the anchor $\rho$ is surjective and ``exact" if $V$ is transitive and $ker(\rho)=im(\rho^*)$, $\rho^*: TX^{\vee} \to V$, leading to the short exact sequence:
\beq
0 \to TX^{\vee} \to V \to TX \to 0.
\eeq
\end{definition}

\subsection{A heterotic Courant algebroid}
In their study of heterotic T-duality, Baraglia and Hekmati define a ``heterotic Courant algebroid" as \cite{Baraglia:2013wua}:
\begin{definition}
A transitive Courant algebroid ${\cal H}$ is defined as a  ``heterotic Courant algebroid" if there exists a principal bundle ${\cal P}$ such that ${\cal A}={\cal H}^{\vee}/TX^{\vee}$ is isomorphic to the Atiyah algebroid of ${\cal P}$ as a quadratic Lie algebroid (i.e. a Lie algebroid with an invariant scalar product):
\begin{align}
& 0 \to {\cal K} \to {\cal H} \to TX \to 0, \\
& 0 \to TX^{\vee} \to {\cal K} \to {\mathfrak g}_{\cal P} \to 0.
\end{align}
\end{definition}
In the above the pairing $\langle , \rangle$ has been used to identify ${\cal H}$ with its dual. In a heterotic theory, the above transitive Courant algebroid exists if ${\cal P}={\cal P}_{total}$ as in \eref{vtot} satisfying the condition $p_1({\cal P}_{total})=0$ (i.e. (\ref{almost_anom}) and (\ref{fiber1}) for $(H_0,F,R)$ satisfying the anomaly cancellation condition) where the first Pontryagin class is defined with respect to a choice of $\langle , \rangle$. For a fixed decomposition
\beq
{\cal H}=TX \oplus \mathfrak{g}_{{\cal P}} \oplus TX^{\vee}~,
\eeq
the anchor, pairing and bracket are given by
\begin{equation}
 \rho(Y,s, \xi)  =  Y, 
\end{equation}
\begin{equation}
\langle (Z, s, \xi), (Y, t, \eta) \rangle  = \frac{1}{2}(i_{Z} \eta+ i_{Y}\xi)+\langle s , t \rangle, 
\end{equation}
\begin{align}
 [Z+s + \xi, Y + t + \eta]_{\cal H} ~=& ~ [Z,Y] + \nabla_{Z}t - \nabla_{Y} s - [s,t]-F(Z,Y)+{\cal L}_{Z}\eta-i_{Y}d\xi+i_{Y}i_{Z}H_0 \nonumber \\
 &~ +2 \langle t, i_{Z} F \rangle -2 \langle s, i_{Y}F \rangle +2 \langle \nabla s, t \rangle,
\end{align}
where $Z,Y \in \Gamma(TX)$, $s,t \in \Gamma (\mathfrak{g}_{\cal P})$, $\xi, \eta \in \Gamma(TX^{\vee})$ and $F$ is the field strength of ${\cal P}_{total}$ above.


\end{document}